\documentclass[10pt]{article}
\usepackage[final]{graphics}
\usepackage{amsmath}
\usepackage{amsfonts,amsbsy}

\def\empile#1\over#2{\mathrel{\mathop{\kern 0pt#1}\limits_{#2}}}

\newcommand{\sll}{\raise.15ex\hbox{$/$}\kern-.43em\hbox{$l$}}
\newcommand{\slepsilon}{\raise.15ex\hbox{$/$}\kern-.53em\hbox{$\epsilon$}}
\newcommand{\slvarepsilon}{\raise.15ex\hbox{$/$}\kern-.53em\hbox{$\varepsilon$}}
\newcommand{\slL}{\raise.15ex\hbox{$/$}\kern-.53em\hbox{$L$}}
\newcommand{\slP}{\raise.15ex\hbox{$/$}\kern-.53em\hbox{$P$}}
\newcommand{\slp}{\raise.1ex\hbox{$/$}\kern-.63em\hbox{$p$}}
\newcommand{\slq}{\raise.1ex\hbox{$/$}\kern-.53em\hbox{$q$}}
\newcommand{\slv}{\raise.1ex\hbox{$/$}\kern-.63em\hbox{$v$}}
\newcommand{\slR}{\raise.15ex\hbox{$/$}\kern-.53em\hbox{$R$}}
\newcommand{\slQ}{\raise.15ex\hbox{$/$}\kern-.53em\hbox{$Q$}}
\newcommand{\slK}{\raise.15ex\hbox{$/$}\kern-.53em\hbox{$K$}}
\newcommand{\slk}{\raise.15ex\hbox{$/$}\kern-.53em\hbox{$k$}}
\newcommand{\slSigma}{\raise.15ex\hbox{$/$}\kern-.53em\hbox{$\Sigma$}}
\newcommand{\slcalP}{\raise.15ex\hbox{$/$}\kern-.63em\hbox{$\cal P$}}
\newcommand{\slA}{\raise.15ex\hbox{$/$}\kern-.73em\hbox{$A$}}
\newcommand{\slbfA}{\raise.15ex\hbox{$/$}\kern-.73em\hbox{${\imb A}$}}
\newcommand{\slpartial}{\raise.15ex\hbox{$/$}\kern-.53em\hbox{$\partial$}}
\newcommand{\sla}{\raise.15ex\hbox{$/$}\kern-.53em\hbox{$a$}}
\newcommand{\slb}{\raise.15ex\hbox{$/$}\kern-.53em\hbox{$b$}}
\newcommand{\slc}{\raise.15ex\hbox{$/$}\kern-.53em\hbox{$c$}}
\newcommand{\slD}{\raise.15ex\hbox{$/$}\kern-.53em\hbox{$D$}}
\newcommand{\slC}{\raise.15ex\hbox{$/$}\kern-.53em\hbox{$C$}}

\def\p{{\boldsymbol p}}
\def\q{{\boldsymbol q}}
\def\l{{\boldsymbol l}}
\def\k{{\boldsymbol k}}

\def\x{{\boldsymbol x}}
\def\y{{\boldsymbol y}}
\def\X{{\boldsymbol X}}
\def\Y{{\boldsymbol Y}}

\def\z{{\boldsymbol z}}
\def\b{{\boldsymbol b}}

\def\u{{\boldsymbol u}}
\def\v{{\boldsymbol v}}

\def\wt{\widetilde}
\def\bs{\boldsymbol}

\def\Aa{A_{_A}}
\def\Apa{A}

\catcode`\@=11

\newcount\@tempcntc
\def\@citex[#1]#2{\if@filesw\immediate\write\@auxout{\string\citation{#2}}\fi
  \@tempcnta\z@\@tempcntb\m@ne\def\@citea{}\@cite{%
        \@for\@citeb:=#2\do%
    {\@ifundefined{b@\@citeb}%
        {\@citeo\@tempcntb\m@ne\@citea%
                \def\@citea{,\penalty\@m\ }{\bf ?}\@warning%
                {Citation `\@citeb' on page \thepage \space undefined}}%
        {\setbox\z@\hbox{\global\@tempcntc0\csname b@\@citeb\endcsname\relax}
     \ifnum\@tempcntc=\z@ \@citeo\@tempcntb\m@ne%
       \@citea\def\@citea{,\penalty\@m}%
       \hbox{\csname b@\@citeb\endcsname}%
     \else%
      \advance\@tempcntb\@ne%
      \ifnum\@tempcntb=\@tempcntc%
      \else\advance\@tempcntb\m@ne\@citeo%
      \@tempcnta\@tempcntc\@tempcntb\@tempcntc\fi\fi}}\@citeo}{#1}}%

\def\@citeo{\ifnum\@tempcnta>\@tempcntb\else\@citea
  \def\@citea{,\penalty\@m}%
  \ifnum\@tempcnta=\@tempcntb\the\@tempcnta\else
   {\advance\@tempcnta\@ne\ifnum\@tempcnta=\@tempcntb \else
\def\@citea{--}\fi
    \advance\@tempcnta\m@ne\the\@tempcnta\@citea\the\@tempcntb}\fi\fi}

\catcode`\@=12

\begin{document}

\thispagestyle{empty}
\title {\bf High energy pA collisions\\
 in the color glass condensate approach\\
 II. Quark production}

\author{Jean-Paul Blaizot$^{(1)}$, Fran\c cois Gelis$^{(1)}$, Raju Venugopalan$^{(2)}$}
\maketitle
\begin{center}
\begin{enumerate}
\item Service de Physique Th\'eorique\footnote{URA 2306 du CNRS.}\\
  B\^at. 774, CEA/DSM/Saclay\\
  91191, Gif-sur-Yvette Cedex, France
\item Physics Department\\
  Brookhaven National Laboratory\\
  Upton, NY 11973, USA
\end{enumerate}
\end{center}

\begin{abstract}
\noindent We compute the production of quark-antiquark pairs in high energy  
collisions between a small and a large projectile, as in
proton-nucleus collisions, in the framework of the Color Glass
Condensate. We derive a general expression for quark pair-production,
which is not $k_\perp$-factorizable. However, $k_\perp$-factorization
is recovered in the limit of large mass pairs or large
quark--anti-quark momenta. Our results are amenable to a simple
interpretation and suggest how multi-parton correlations at small $x$
can be quantified in high-energy proton/deuteron-nucleus collisions.

\end{abstract}

\section{Introduction}
In an accompanying paper~\cite{BlaizGV1}, heretofore referred to as I,
we developed the Color Glass Condensate (CGC)
framework~\cite{IancuV1,IancuLM3,Muell4} to address the case where the
parton densities in the projectile are small while those of the target
are large. This may be applied for instance to the
proton/deuteron-nucleus collisions that are currently being studied at
Brookhaven's Relativistic Heavy Ion Collider (RHIC) and will be
further studied at the Large Hadron Collider at CERN in the near
future. In our previous paper \cite{BlaizGV1}, we solved the
Yang-Mills equations, for two light-cone sources, to first order in
the density of the dilute projectile and to all orders in the density
of the target. Our computations were performed in the Lorenz/covariant
gauge $\partial_\mu A^\mu=0$. Paper I focused on developing a general
formalism for gluon production at high energies and on
phenomenological consequences such as the Cronin effect.  In I, we
derived a compact expression of the gauge fields produced in
proton-nucleus collisions. In this paper, we will use this result to
study the production of quark--anti-quark pairs in high energy
proton-nucleus collisions\footnote{Note that this problem has also
  been investigated numerically by Dietrich using a toy model for the
  background classical color field \cite{Dietr1,Dietr2}.}.

Quark pair production in high energy hadronic collisions has
previously been studied in the frameworks of collinear
factorization~\cite{NasonDE1,NasonDE2,FrixiMNR1} and
$k_\perp$-fac\-to\-ri\-za\-tion~\cite{ColliE1,CatanCH1,LevinRSS1}. A
comprehensive list of references for the particular applications of
these formalisms to proton-nucleus collisions can be found in
ref.~\cite{Accara2}. At high energies, where the intrinsic transverse
momenta of the partons is not negligible and may be on the order of
the saturation scale~\cite{LevinRSS1}, $k_\perp\sim Q_s$,
$k_\perp$-factorization is a convenient formalism to discuss particle
production. Phenomenological studies for hadron colliders have been
discussed in refs.~\cite{HagleKSST1,RyskiSS1} and for heavy-ion
collisions in ref.~\cite{KharzT1}. $k_\perp$-factorization was proven
for large mass pair-production at small $x$ in
refs.~\cite{ColliE1,CatanCH1,LevinRSS1}.

The issue of $k_\perp$-factorization for quark production in the CGC
approach was addressed by two of us in~\cite{GelisV1}. We showed that,
in the CGC language, the $k_\perp$-fac\-to\-ri\-za\-tion of Collins and
Ellis~\cite{ColliE1} is recovered exactly\footnote{The results of
Collins and Ellis are equivalent to those of Catani, Ciafaloni and
Hautmann~\cite{CatanCH1}.} when the parton densities of the projectile
and the target are not too large, namely, when
$\rho_p/k_\perp^2,\rho_{_{A}}/k_\perp^2 \ll 1$. Here $\rho_p$,
$\rho_{_{A}}$ denote the number of parton sources per unit area in the
projectile and target respectively and $k_\perp$ is a typical momentum
of the produced parton. Our results also suggested however that
$k_\perp$-factorization is not robust and would fail when either or
both $\rho_p,\rho_{_{A}} \sim k_\perp^2$. High energy proton-nucleus
collisions are ideal to test this conjecture explicitly since the
proton source may be taken to be dilute ($\rho_p/k_\perp^2 \ll1$)
while the nuclear target is dense ($\rho_{_{A}}/k_\perp^2 \sim
1$). For gluon production, we (and others previously) showed that
$k_\perp$-factorization remains valid even in proton-nucleus
collisions, although one has to use an ``unintegrated nuclear gluon
distribution'' which is not the canonical one. However, we will show
in this paper that $k_\perp$-factorization fails unequivocally for
pair production in pA-collisions. Nevertheless, our final result
(eq.~\ref{eq:cross-section}) can be expressed in terms of a small
number of distribution functions which have  a simple interpretation.

Unlike the unintegrated distribution discussed in the case of gluon
production, these distributions cannot all be expressed in terms of
2-point correlators of adjoint Wilson lines.  In addition to a term
containing the correlator $\big<U(\x_\perp) U^\dagger(\y_\perp)\big>$,
one has a term proportional to $\big<{\wt U}(\x_\perp) t^a {\wt
U}^\dagger(\y_\perp)t^b U^{ba}(\x_\perp^\prime)\big>$, i.e. a product of
two Wilson lines in the fundamental representation (denoted ${\wt U}$)
and an adjoint Wilson line, and a third term proportional to the
product of four fundamental Wilson lines: $\big<{\wt U}(\x_\perp)t^a
{\wt U}^\dagger(\y_\perp){\wt U}^\dagger(\y_\perp^\prime)t^a {\wt
U}^\dagger(\x_\perp^\prime)\big>$.  The transverse coordinates here
represent those of the quark and the anti-quark in the pair production
amplitude and in the complex conjugate amplitude. For very large mass
pairs and for large momenta of the pair, the transverse separation of
the pair can be neglected, and one can show, through an identity, that
the $k_\perp$-factorized result in terms of the two point function of
adjoint Wilson lines is recovered.

All the above Wilson lines are computed in the presence of the
classical color field of the nucleus and therefore depend on the
source $\rho_{_{A}}$. The brackets $\left<\cdots \right>$ here denote
the weighted quantum mechanical average of an operator in the
background field of the proton and the nucleus, which can be expressed
as
\begin{eqnarray}
\langle O\rangle
=\int [D\rho_p][D\rho_{_{A}}]
W_p[x_0^p,\rho_p]W_{_{A}}[{x_0^{_A}},\rho_{_{A}}]
\, O[\rho_p,\rho_{_{A}}] \, .
\label{eq:average}
\end{eqnarray}
This averaging procedure is essential in order to restore gauge
invariance since $O[\rho_p,\rho_{_{A}}]$ is computed in a particular
gauge.  More importantly, it is also through $W_p$ and $W_{_{A}}$ that
quantum effects, due to evolution of the light cone wave functions of
the target and projectile with $x$, are incorporated in the study of
pair production at high energies.  The arguments $x_0^p$ and
$x_0^{_A}$ denote the scale in $x$ separating the large-$x$ static
sources from the small-$x$ dynamical fields. In the
McLerran-Venugopalan model, the functional $W_{_{A}}$ that describes
the distribution of color sources in the nucleus is a Gaussian in
$\rho_{_{A}}$ \cite{McLerV1,McLerV2,McLerV3,McLerV4}.  In general,
this Gaussian is best interpreted as the initial condition for a
non-trivial evolution of $W_{_{A}}[x_0^{_A},\rho_{_{A}}]$ with
$x_0^{_A}$. This evolution is described by a Wilson renormalization
group equation (often called the JIMWLK
equation~\cite{JalilKLW1,JalilKLW2,JalilKLW3,JalilKLW4,KovneM1,KovneMW3,Balit1,Kovch1,Kovch3,JalilKMW1,IancuLM1,IancuLM2}).
This evolution equation has been solved recently numerically by
Rummukainen and Weigert~\cite{RummuW1}. One is therefore in a position
to make predictions for high energy pair-production in hadronic
collisions that are sensitive to not only the ``unintegrated gluon
distribution'' but to more subtle correlations among partons in
hadrons at high energies. Conversely, comparisons with pair production
experiments at high energies can test, with a high degree of
precision, whether the JIMWLK equations, or equivalently the Color
Glass Condensate, is the right effective theory of high energy QCD.

This paper is organized as follows. In section \ref{sec:field}, we
briefly review the solution to the Yang-Mills equations derived in I
and re-express it in a way convenient for our discussion. We next
proceed to the computation of the quark pair production amplitude in
section \ref{sec:F-amplitude}. This amplitude consists of two sets of
terms which we name ``regular'' and ``singular'' terms. The former
correspond to the following cases: a) the gluon produces the pair
before interacting with the nucleus -- the pair therefore scatters on
the nucleus on its way out, and b) the gluon interacts with the
nucleus and produces the pair after the encounter.  The singular case
corresponds to the situation where the pair is produced inside the
nucleus and re-interacts on the way out.  Naively, one would not
expect this term to contribute at very high energies, due to the
brevity of the encounter. Nevertheless, in the Lorenz gauge that we
choose to work in, such a term does exist and is important in
obtaining our final result.  Adding both sets of terms together, we
obtain the time-ordered amplitude for pair production. For
completeness, and to confirm our result for the time ordered
amplitude, we also compute the retarded pair production amplitude,
which, to this order, carries an identical physical content albeit a
different interpretation. The computation is presented in appendix
\ref{sec:ret-amp}.  In section \ref{eq:cs}, we compute the pair
production cross-section. Our final result
(eq.~\ref{eq:cross-section}) is not $k_\perp$-factorizable in the same
manner as the result for the gluon production cross-section is.  The
various terms in our result however have a simple interpretation and
for large mass or large momentum pairs, one does recover the
$k_\perp$-factorized result - as expected from our derivation
in~\cite{GelisV1}.  We finally address the case of the single quark
production cross-section in section \ref{sec:single-quark}.  Single
quark distributions were recently studied by Tuchin \cite{Tuchi1} in
the context of a Glauber model of independent scatterings. In contrast
to our momentum space approach, Tuchin's results are formulated
entirely in coordinate space and therefore do not address the issue
of $k_\perp$-factorization. Further, our results are more general,
being valid for non-Gaussian correlations as well.  For the case of a
Gaussian distribution of color sources, we provide in appendix
\ref{sec:4-point-corr} explicit expressions for all the correlators
that appear in our results for both pair production and inclusive
single quark production.  We end with a summary of our results and an
outlook on future work on these questions.

\section{The Yang-Mills gauge field produced in  pA collisions}
\label{sec:field}
Our first step in calculating the pair production amplitude is to find
solutions of the classical Yang-Mills equations in the presence of
proton and nuclear sources of color charge. Since we want to compute
quark production at the lowest order in the source $\rho_p$ that
describes the proton and to all orders in the source $\rho_{_{A}}$
that describes the nucleus, we need to compute the gauge field up to
the same order. The Yang-Mills equations, to this required order, were
solved in covariant gauge in I. We will here just quote the results
and re-express them in a manner convenient for the computation of the
pair production amplitude in the next section.  Following I, we denote
$A_{_{A}}^\mu$ the field of the nucleus alone, $A_p^\mu$ the field of
the proton alone and $A^\mu$ (without any subscript) the total gauge
field at order $1$ in the proton source and to all orders in the
nuclear source.

The gauge field created by the nucleus alone is given in coordinate
space by the expression
\begin{equation}
\Aa^\mu(x)=-g\delta^{\mu-}\delta(x^+)\frac{1}{{\bs\nabla}_\perp^2}
\rho_{_{A}}(\x_\perp)\; .
\end{equation}
Its  Fourier transform, which will appear repeatedly as we proceed, is 
\begin{equation}
A_{_A}^\mu(q)=2\pi g\delta^{\mu-}\delta(q^+)
\frac{{{\rho}_{A}}(\q_\perp)}{q_\perp^2}\; .
\label{eq:field-0}
\end{equation}
Note that this field is linear in the source $\rho_{_{A}}$: it is
indeed a well known fact that, in covariant gauge, the classical color
field created by a single projectile, no matter how dense, is linear
in the classical color source. We denote as $\Apa^\mu$ the first
correction to this field due to the proton source, which is of order
one in $\rho_p$. This correction was derived in the covariant gauge in
I (see section 3.5 of I).

Our expression for $\Apa^\mu$ has both singular terms (proportional to
$\delta(x^+)$) and regular terms. It will be convenient for our later
discussion to group them accordingly, and express $\Apa^\mu$ as
\begin{eqnarray}
A^\mu(q)= A_{{\rm reg}}^\mu(q)
+\delta^{\mu -} A_{\rm sing}^-(q)\; .
\end{eqnarray}
Our expression for $ A_{{\rm reg}}^\mu$ is 
 \begin{eqnarray}
 A_{{\rm reg}}^\mu(q)&=& A_p^\mu(q)
\nonumber\\
&+&
\frac{ig}{q^2+iq^+\epsilon}\int\frac{d^2\k_{1\perp}}{(2\pi)^2}
\bigg\{
C_{_{U}}^\mu(q,\k_{1\perp})\, 
\left[U(\k_{2\perp})-(2\pi)^2\delta(\k_{2\perp})\right]
\nonumber\\
&&\qquad\qquad\;\;
+
C_{_{V},{\rm reg}}^\mu(q)\, 
\left[V(\k_{2\perp})-(2\pi)^2\delta(\k_{2\perp})\right]
\bigg\}\frac{{\rho_p}(\k_{1\perp})}{k_{1\perp}^2}\; .
\label{eq:field-1}
\end{eqnarray}
In this formula, the first term is the color field of the proton
alone, given by eq.~(\ref{eq:field-0}) with $\rho_{_{A}}$ replaced by
$\rho_p$. In the second term, $\k_{1\perp}$ is the momentum coming
from the proton and $\k_{2\perp}$, defined as $\k_{2\perp}\equiv
\q_\perp-\k_{1\perp}$, is the momentum coming from the nucleus. The
4-vectors $C_{_{U}}^\mu(q,\k_{1\perp})$ and $C_{_{V},{\rm
reg}}^\mu(q)$ are given by the following relations:
\begin{eqnarray}
&& 
C_{_{U}}^+(q,\k_{1\perp})\equiv -\frac{k_{1\perp}^2}{q^-+i\epsilon}\;,\; 
C_{_{U}}^-(q,\k_{1\perp})\equiv \frac{k_{2\perp}^2-q_\perp^2}{q^+}\;,\;
C_{_{U}}^i(q,\k_{1\perp})\equiv -2 k_1^i\; ,
\nonumber\\
&& 
C_{_{V},{\rm reg}}^+(q)\equiv 2q^+ \quad,\quad 
C_{_{V},{\rm reg}}^-(q)\equiv 2q^--\frac{q^2}{q^+}\quad,\quad 
C_{_{V},{\rm reg}}^i(q)\equiv 2 q^i
\; .
\end{eqnarray}
$U$ and $V$ are the Fourier transforms of
Wilson lines in the adjoint representation of $SU(N)$:
\begin{eqnarray}
U(\k_{\perp})=\int d^2\x_\perp e^{i\k_{\perp}\cdot\x_\perp}
U(\x_\perp)\; ,\;\;\;
V(\k_{\perp})=\int d^2\x_\perp e^{i\k_{\perp}\cdot\x_\perp}
V(\x_\perp)\; ,
\end{eqnarray}
with
\begin{eqnarray}
&&
U(\x_\perp)\equiv {\cal P}_+ \exp\left[ig\int_{-\infty}^{+\infty}
dz^+ A_{_A}^-(z^+,\x_\perp)\cdot T
\right]\; ,\nonumber\\
&&V(\x_\perp)\equiv {\cal P}_+ \exp\left[i\frac{g}{2}\int_{-\infty}^{+\infty}
dz^+ A_{_A}^-(z^+,\x_\perp)\cdot T
\right]\; ,
\end{eqnarray}
where the $T^a$ are the generators of the adjoint representation of
$SU(N)$ and ${\cal P}_+$ denotes a ``time ordering'' along the $z^+$
axis. The subscript `${\rm reg}$' on $C_{_{V}}^\mu$ indicates that the
corresponding term of $A^\mu$ does not contain any $\delta(x^+)$ when
expressed in coordinate space. The relationship\footnote{The terms
responsible for a $\delta(x^+)$ in $A^\mu(x)$ are those for which
$C_{_{U,V}}^\mu/q^2$ does not go to zero when $q^-\to\infty$. It is
immediate to verify that this happens only for
$C_{_{V}}^-$. $C_{_{V},{\rm reg}}^-$ is obtained from $C_{_{V}}^-$ the
piece responsible for this non-zero limit.} between $C_{_{V},{\rm
reg}}^\mu(q)$ and the vector $C_{_{V}}^\mu(q)$ introduced in the
section 3.5 of I is\footnote{$C_{_{U}}$ and $C_{_{V}}$ are related to
the well known Lipatov vertex~\cite{Lipat1,Duca1} via the relation
$C_{_{L}}^\mu = C_{_{U}}^\mu + \frac{1}{2}\, C_{_{V}}^\mu$. The
properties of these are explored further in paper I.}:
\begin{equation}
C_{_{V},{\rm reg}}^\mu(q)=C_{_{V}}^\mu(q)+\delta^{\mu -}\frac{q^2}{q^+}\; .
\end{equation}
The ``singular'' term reads:
\begin{eqnarray}
A_{\rm sing}^-(q)\equiv
-\frac{ig}{q^+}\int \frac{d^2\k_{1\perp}}{(2\pi)^2}
\left[V(\k_{2\perp})-(2\pi)^2\delta(\k_{2\perp})\right]
\frac{{\rho_p}(\k_{1\perp})}{k_{1\perp}^2}\; .
\end{eqnarray}
It will in fact be more convenient later to have this part of the
field in coordinate space:
\begin{eqnarray}
{A}_{\rm sing}^-(x)=
i\frac{g^2}{2}(A_{_A}^-(x)\cdot T)
V(x^+,-\infty;\x_\perp)
\theta(x^-)\frac{1}{{\bs\nabla}_\perp^2}\rho_p(\x_\perp)\; ,
\label{eq:sing-field}
\end{eqnarray}
where $V(x^+,-\infty;\x_\perp)$ denotes an incomplete Wilson
line:
\begin{eqnarray}
V(x^+,-\infty;\x_\perp)\equiv {\cal P}_+ 
\exp\left[i\frac{g}{2}\int_{-\infty}^{x^+}
dz^+ A_{_A}^-(z^+,\x_\perp)\cdot T
\right]\; .
\end{eqnarray}
In the case of gluon production I, we have seen that the
Wilson line $V$ drops out of the gluon production amplitude. This
result was to be expected from the fact that there are gauges in which
such a Wilson line, characterized by an unusual factor $1/2$ in the
exponent, does not appear at all. Naturally, we expect the same for
quark-antiquark production, and one of our tasks in the next section
will be to exhibit the mechanism by which this Wilson line drops out
of the pair production amplitude.

\section{Pair production amplitude}
\label{sec:F-amplitude}
\subsection{Generalities}
Once we know the time-dependent classical gauge field created in the
collision of the two projectiles, it is straightforward to calculate
the production of quark-antiquark pairs in the collision. One can
simply forget about the projectiles themselves, and consider only the
classical field. The probability to produce exactly one $q\bar{q}$
pair\footnote{If one uses instead the retarded quark-propagator, one
obtains the average number $\overline{N}_{q\bar{q}}$ of produced pairs
in the collision. At leading order in $\rho_p$, at most one pair can
be produced per collision, so that one expects
$P_1=\overline{N}_{q\bar{q}}$. We check this property explicitly in
 appendix \ref{sec:ret-amp}.}  in the collision can be expressed in
terms of the time-ordered quark propagator in the presence of the
classical field\footnote{This formula for $P_1$ is correct at leading
order in $\rho_p$, but is incomplete in general.  Indeed, it should
also contain the prefactor $\left|\left<0_{\rm in}|0_{\rm
out}\right>\right|^2$, which is mandatory for unitarity to be
preserved (see \cite{BaltzGMP1} for more details).  This prefactor
only affects higher order corrections in $\rho_p$.} as 
\begin{equation}
P_1[\rho_p,\rho_{_A}]=
\int\frac{d^3\q}{(2\pi)^3 2E_\q}
\int\frac{d^3\p}{(2\pi)^3 2E_\p}
\left|{\cal M}_{_{F}}(\q,\p)\right|^2\; ,
\label{eq:P1-def}
\end{equation}
where the argument $[\rho_p,\rho_{_A}]$ indicates that this is the
production probability in one particular configuration of the color
sources. In order to turn this probability into a cross-section, one
must average over the initial classical sources $\rho_p$ and
$\rho_{_{A}}$ with the weight $W_p[\rho_p]W_{_{A}}[\rho_{_{A}}]$ and one
must integrate over all the impact parameters $\b$, to obtain,
\begin{equation}
\sigma_1=\int d^2\b \int[D\rho_p][D\rho_{_{A}}]W_p[\rho_p]W_{_{A}}[\rho_{_{A}}]
P_1[\rho_p,\rho_{_A}]\; .
\end{equation}
In eq.~(\ref{eq:P1-def}), the time-ordered amplitude ${\cal M}_{_{F}}$
is related to the quark propagator as follows:
\begin{equation}
{\cal M}_{_{F}}(\q,\p)\equiv \overline{u}(\q)
{\cal T}_{_{F}}(q,-p)
 v(\p)\; ,
\end{equation}
where ${\cal T}_{_{F}}(q,-p)$ is the interacting part of the Feynman
quark propagator in the presence of the classical field, with its external legs amputated. The arguments $q$ and $-p$ are respectively the
outgoing and the incoming 4-momenta on this quark propagator.

\subsection{Regular terms}
Since we want the pair production amplitude to first order in the
source $\rho_p$ and to all orders in $\rho_{_{A}}$, we must insert the
total field $\Apa^\mu$ exactly once on the quark propagator, and we
can insert the field $\Aa^\mu$ an arbitrary number of times since it
does not change the power counting in $\rho_p$. At this point, it is
convenient to study separately the terms in $\Apa^\mu$ which do not
contain a $\delta(x^+)$ from those that contain a $\delta(x^+)$.

\begin{figure}[htbp]
\begin{center}
\resizebox*{!}{4cm}{\includegraphics{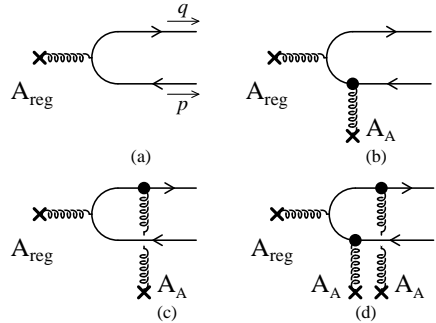}}
\end{center}
\caption{\label{fig:reg-diagrams} Regular terms in the time-ordered
pair production amplitude. The gluon line terminated by a cross
denotes a classical field insertion. The black dot denotes multiple
insertions of the field $\Aa^\mu$ (with at least one insertion).}
\end{figure}

Let us begin with the former. The terms that do not contain a
$\delta(x^+)$ are the ones in eq.~(\ref{eq:field-1}) that we called
$\Apa_{\rm reg}^\mu$. Physically, these are terms for which the pair
is produced outside the nucleus; more precisely, the point where
$\Apa^\mu$ is inserted on the quark line must be outside the
nucleus. Regular terms are represented by the four diagrams of figure
\ref{fig:reg-diagrams}.
In order to calculate these diagrams, we need the expression for the
multiple insertions of the field $\Aa^\mu$ on the time-ordered quark
propagator, denoted by a black dot in the diagrams. This reads
\cite{BaltzGMP1,GelisP1}:
\setbox1=%
\hbox to 1.3cm{\resizebox*{1.3cm}{!}{\includegraphics{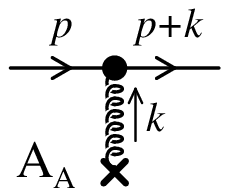}}}
\begin{equation}
\raise -4mm\box1=2\pi\delta(k^+)\gamma^+\,{\rm sign}(p^+)
\int d^2\x_\perp e^{i\k_\perp\cdot\x_\perp}
\left[
{\wt U}^{{\rm sign}(p^+)}(\x_\perp)-1
\right]\; ,
\label{eq:F-scatt}
\end{equation}
where ${\wt U}$ is a Wilson line in the fundamental representation of
$SU(N)$, defined by
\begin{equation}
{\wt U}(\x_\perp)\equiv {\cal P}_+
\exp\left[ig\int_{-\infty}^{+\infty} dz^+
A_{_A}^-(z^+,\x_\perp)\cdot t\right]\; .
\end{equation}
The $t^a$ are the generators of the fundamental representation
of $SU(N)$. The contributions of the diagrams of figure
\ref{fig:reg-diagrams} to the time-ordered pair production amplitude
are obtained  using the free Feynman (time-ordered) quark
propagator in the intermediate quark lines:
\begin{equation}
S_{_{F}}^0(p)\equiv i\frac{\slp+m}{p^2-m^2+i\epsilon}\; .
\label{eq:Sf}
\end{equation}
The contribution of the diagram (a) to ${\cal M}_{_{F}}$ is simply
given by
\begin{equation}
{\cal M}_{_{F}}^{(a)}(\q,\p)=
\big[-ig A_{\rm reg}^{\mu a}(p+q)\big]
\big[\overline{u}(\q)\gamma_\mu t^a v(\p)\big]\; .
\end{equation}
Since the first term of eq.~(\ref{eq:field-1}) brings a
$\delta(p^-+q^-)$ with it, this term cannot contribute to the
production of a pair on shell, for which we must have $p^->0$ and
$q^->0$. (This first term would correspond to the production of a pair
from the proton alone, which is kinematically forbidden). Using also
$\overline{u}(\q)(\slp+\slq)v(\p)=0$, we can write the contribution
from diagram (a) more explicitly as
\begin{eqnarray}
&&{\cal M}_{_{F}}^{(a)}(\q,\p)=g^2\int\frac{d^2\k_{1\perp}}{(2\pi)^2}
\frac{{\rho}_{p,a}(\k_{1\perp})}{k_{1\perp}^2}
\int d^2\x_\perp
e^{i(\p_\perp+\q_\perp-\k_{1\perp})\cdot\x_\perp}
\nonumber\\
&&\!\!\!\!\!\!\!\!
\times
\Big\{
\frac{\overline{u}(\q)\slC_{_{U}}(p+q,\k_{1\perp})t^b v(\p)}{(p+q)^2}
\left[U(\x_\perp)\!-\!1\right]_{ba}
\!-\!\frac{\overline{u}(\q)\gamma^+ t^b v(\p)}{p^++q^+}
\left[V(\x_\perp)\!-\!1\right]_{ba}\!\!
\Big\}\; .\nonumber\\
&&
\label{eq:Mf-a}
\end{eqnarray}
The contribution of the diagram (b) reads:
\begin{eqnarray}
&&{\cal M}_{_{F}}^{(b)}(\q,\p)=-\int\frac{d^4k}{(2\pi)^4}
2\pi\delta(k^+)
\big[-ig A_{\rm reg}^{\mu a}(p+q-k)\big]
\int d^2\x_\perp e^{i\k_\perp\cdot\x_\perp}
\nonumber\\
&&\qquad\qquad\qquad\qquad\times
\overline{u}(\q)
\gamma_\mu t^a
S_{_{F}}^0(-p+k)
\gamma^+
\left[{\wt U}^\dagger(\x_\perp)-1\right]v(\p)\; ,
\end{eqnarray}
where we have already used the fact that the quark propagator on which
the eikonal interaction with the nucleus is inserted carries a
negative energy $-p^+<0$. The $k^+$ integral is trivial because of the
$\delta(k^+)$. The $k^-$ integral can be performed by using the
theorem of residues. The propagator $S_{_{F}}^0(-p+k)$ has a pole
above the real axis in the $k^-$ plane. So do the denominators
$(p+q-k)^2$ (the prefactor in the second term of
eq.~(\ref{eq:field-1})) and $p^-+q^--k^-+i\epsilon$ (in
$C_{_{U}}^+$). Therefore, we see that the second line in the field
$A_{\rm reg}^\mu$ cannot contribute because it has all its poles on the
same side of the real axis\footnote{For this conclusion to be valid,
it was crucial to have isolated the singular terms in the gauge field
from the regular ones. Indeed, the singular terms have a large $k^-$
behavior which prevent one from closing the integration axis by the
contour at infinity. }. The only term from $A_{\rm reg}^\mu$ that
contributes is the first one, namely, the field of the proton
alone. Finally, using the variable $\k_{1\perp}\equiv
\p_\perp+\q_\perp-\k_\perp$ instead of $\k_\perp$, we have
\begin{eqnarray}
&&{\cal M}_{_{F}}^{(b)}(\q,\p)=g^2\int\frac{d^2\k_{1\perp}}{(2\pi)^2}
\frac{\rho_{p,a}(\k_{1\perp})}{k_{1\perp}^2}
\int d^2\x_\perp
e^{i(\p_\perp+\q_\perp-\k_{1\perp})\cdot\x_\perp}
\nonumber\\
&&\qquad\qquad
\times
\frac
{\overline{u}(\q)\gamma^- t^a (\slq-\slk_1+m)\gamma^+
[{\wt U}^\dagger(\x_\perp)-1]v(\p)}
{2p^+q^-+(\q_\perp-\k_{1\perp})^2+m^2}\; .
\label{eq:Mf-b}
\end{eqnarray}
Similarly for the diagram (c), the poles in the variable $k^-$ are
such that only the first term in $A_{\rm reg}^\mu$
contributes. We obtain
\begin{eqnarray}
&&{\cal M}_{_{F}}^{(c)}(\q,\p)=-g^2\int\frac{d^2\k_{1\perp}}{(2\pi)^2}
\frac{\rho_{p,a}(\k_{1\perp})}{k_{1\perp}^2}
\int d^2\x_\perp
e^{i(\p_\perp+\q_\perp-\k_{1\perp})\cdot\x_\perp}
\nonumber\\
&&\qquad\qquad
\times
\frac
{\overline{u}(\q)\gamma^+ [{\wt U}(\x_\perp)-1] 
(-\slp+\slk_1+m)\gamma^- t^a
v(\p)}
{2p^-q^++(\p_\perp-\k_{1\perp})^2+m^2}\; .
\label{eq:Mf-c}
\end{eqnarray}
By the same method, we can write the contribution of the diagram (d)
as follows:
\begin{eqnarray}
&&\!\!\!\!\!\!\!\!\!\!\!\!
{\cal M}_{_{F}}^{(d)}(\q,\p)\!=\!g^2\!\!\int\!\frac{d^2\k_{1\perp}}{(2\pi)^2}
\frac{d^2\k_\perp}{(2\pi)^2}
\frac{\rho_{p,a}(\k_{1\perp})}{k_{1\perp}^2}
\!\int\!\! d^2\x_\perp d^2\y_\perp
e^{i\k_\perp\cdot\x_\perp}
e^{i(\p_\perp\!+\!\q_\perp\!-\!\k_\perp\!-\!\k_{1\perp})\cdot\y_\perp}
\nonumber\\
&&\!\!\!\!\!\!
\times
\frac
{\overline{u}(\q)\gamma^+ [{\wt U}(\x_\perp)\!-\!1] 
(\slq-\slk+m)\gamma^- t^a (\slq-\slk-\slk_1+m)\gamma^+
[{\wt U}^\dagger(\y_\perp)\!-\!1]
v(\p)}
{2p^+[(\q_\perp-\k_\perp)^2+m^2]+2q^+[(\q_\perp-\k_\perp-\k_{1\perp})^2+m^2]}\; .\nonumber\\
&&
\label{eq:Mf-d}
\end{eqnarray}
In this expression, $\x_\perp$ and $\y_\perp$ are the transverse
coordinates of the quark and the antiquark respectively, $\k_{1\perp}$
is the momentum flowing from the proton, $\k_\perp$ the momentum
flowing from the nucleus to the quark line, and
$\p_\perp+\q_\perp-\k_\perp-\k_{1\perp}$ the momentum flowing from the
nucleus on the antiquark line.

One can see at this point that the sum of the regular
terms\footnote{In fact, only the diagram (a) contains the Wilson line
$V$. For the diagrams (b), (c) and (d), only the term $A_p^\mu$ of the
total field $A^\mu$ contributes, for kinematical reasons.}  contains
the Wilson line $V$. As we shall see shortly, it is cancelled by the
contribution coming from the singular terms.

\subsection{Singular terms}
\label{sec:F-singular-term}
\begin{figure}[htbp]
\begin{center}
\resizebox*{!}{3cm}{\includegraphics{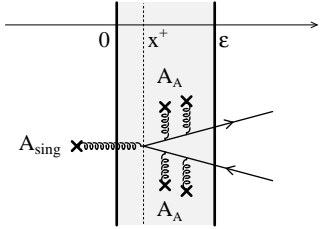}}
\end{center}
\caption{\label{fig:sing-diagram} Singular term in the time-ordered
pair production amplitude. The gluon line terminated by a cross
denotes a classical field insertion.}
\end{figure}

We must now evaluate the contribution to the pair production amplitude
coming from the field $\Apa_{\rm sing}^\mu$. Since this field is
proportional to $\delta(x^+)$ in coordinate space, its contribution is
a term where the pair is produced inside the nucleus. Technically,
this means that in order to correctly compute this term, we must go to
coordinate space and regularize the $\delta(x^+)$ by giving a small
width to the nucleus,
\begin{equation}
\delta(x^+)\to \delta_\epsilon(x^+)\; ,
\end{equation}
where $\delta_\epsilon(x^+)$ is a regular function (whose support is
$[0,\epsilon]$), which becomes a $\delta(x^+)$ when $\epsilon$ goes to
zero. We need not specify this function further because the results do
not depend on the precise choice of the regularization. The singular
contribution is depicted in figure \ref{fig:sing-diagram}. The field
$\Apa_{\rm sing}^\mu$ is inserted on the quark line at the `time'
$x^+$; the quark and the anti-quark then rescatter on the field
$\Aa^\mu$ of the nucleus between $x^+$ and $\epsilon$.
This contribution to the amplitude can be written  in
coordinate space as, 
\begin{eqnarray}
&&
{\cal M}_{_{F}}^{\rm sing}(\q,\p)=\int d^4x
e^{i(p+q)\cdot x}
\overline{u}(\q)
{\wt U}(+\infty,x^+;\x_\perp)
\nonumber\\
&&\qquad\qquad
\times
\big[ig\gamma^+ t^a \Apa_{\rm sing}^{-a}(x)\big]
{\wt U}^\dagger(+\infty,x^+;\x_\perp)v(\p)\; ,
\end{eqnarray}
where we have introduced the incomplete Wilson line
\begin{equation}
{\wt U}(+\infty,x^+;\x_\perp)\equiv
{\cal P}_+
\exp\left[ig\int_{x^+}^{+\infty}
dz^+ A_{_A}^-(z^+,\x_\perp)\cdot t
\right]\; .
\end{equation}
Note that here the fundamental Wilson line and its complex conjugate
are evaluated at the same point $\x_\perp$. This is because in the
limit $\epsilon\rightarrow 0$, at the end of the interaction, the
components of the pair will not have had sufficient time to separate
from each other. Note further that the four possible contributions
(analogue to the four regular terms) are already summed here by
replacing ${\wt U} -1$ by ${\wt U}$ and ${\wt U}^\dagger-1$ by ${\wt
U}^\dagger$.

In order to simplify this expression, we need the following algebraic
identity:
\begin{equation}
{\wt U}(+\infty,x^+;\x_\perp) t^a {\wt U}^\dagger(+\infty,x^+;\x_\perp)
=t^b U^{ba}(+\infty,x^+;\x_\perp)\; .
\label{eq:wilson-alg-1}
\end{equation}
Using also the explicit form eq.~(\ref{eq:sing-field}) of the singular
field, we can write the singular contribution to the amplitude as\footnote{We have used the fact that the support of
the $x^+$ integral is $0< x^+<\epsilon$ in order to approximate
$\exp(i(p^-+q^-)x^+)\approx 1$.},
\begin{eqnarray}
&&{\cal M}_{_{F}}^{\rm sing}(\q,\p)=
ig^2\int dx^- d^2\x_\perp 
e^{i(p^++q^+)x^-}e^{-i(\p_\perp+\q_\perp)\cdot\x_\perp}
\nonumber\\
&&\qquad\times
\left\{
i\frac{g}{2}
\int_{-\infty}^{+\infty}
dx^+\big[
U(+\infty,x^+;\x_\perp)
(\Aa^-(x)\cdot T)
V(x^+,-\infty;\x_\perp)
\big]_{ab}
\right\}
\nonumber\\
&&\qquad\times
\overline{u}(\q)\gamma^+ t^a 
\theta(x^-)\frac{1}{{\bs\nabla}_\perp^2}\rho_{p,b}(\x_\perp)
v(\p)\; .
\label{eq:Mf-sing-tmp}
\end{eqnarray}
Finally we need the formula
\begin{equation}
i\frac{g}{2}\!\!
\int\limits_{-\infty}^{+\infty}\!\!
dx^+
U(+\infty,x^+;\x_\perp)
(\Aa^-(x)\!\cdot\! T)
V(x^+,-\infty;\x_\perp)
\!=\!U(\x_\perp)\!-\!V(\x_\perp)\; ,
\label{eq:wilson-alg-2}
\end{equation}
which we encountered previously in I. The remaining Fourier integral
in eq.~(\ref{eq:Mf-sing-tmp}) can be performed explicitly, and we
obtain
\begin{eqnarray}
&&{\cal M}_{_{F}}^{\rm sing}(\q,\p)=g^2\int\frac{d^2\k_{1\perp}}{(2\pi)^2}
\frac{\rho_{p,a}(\k_{1\perp})}{k_{1\perp}^2}
\int d^2\x_\perp
e^{i(\p_\perp+\q_\perp-\k_{1\perp})\cdot\x_\perp}
\nonumber\\
&&\qquad\qquad\qquad
\times
\frac{\overline{u}(\q)\gamma^+ t^b v(\p)}{p^++q^+}
\big[
V(\x_\perp)-U(\x_\perp)
\big]_{ba}\; .
\label{eq:Mf-sing}
\end{eqnarray}
We see that, as anticipated, the term in $V$ in ${\cal M}_{_{F}}^{\rm
sing}$ cancels the term in $V$ we found previously in ${\cal
M}_{_{F}}^{(a)}$.

\subsection{Complete time-ordered amplitude}
We must now combine the terms ${\cal M}_{_{F}}^{(a,b,c,d)}$ and ${\cal
M}_{_{F}}^{\rm sing}$.  In order to do so, we shall use the following
identities:
\begin{eqnarray}
&&\overline{u}(\q)\gamma^-=\frac{1}{2q^+}\overline{u}(\q)\gamma^+
(\slq+m)\gamma^-\; ,\nonumber\\
&&\gamma^- v(\p)=\frac{1}{2p^+}\gamma^-(\slp-m)\gamma^+ v(\p)\; ,
\end{eqnarray}
which are simple consequences of the Dirac equation obeyed by the free
spinors. The second trick is to introduce dummy variables $\k_\perp$
and $\y_\perp$ via the identity 
\begin{equation}
1=\int\frac{d^2\k_\perp}{(2\pi)^2}\int d^2\y_\perp\; 
e^{i\k_\perp\cdot\y_\perp}\; ,
\end{equation}
so that all the terms have the same integration variables. Putting all
these terms together, we obtain the following expression for the complete
amplitude\footnote{In order to obtain this expression, we have also
used a cancellation between the $-t^a$ terms in ${\wt U}t^a {\wt
U}^\dagger-t^a$ and in $t^b[U-1]_{ba}$.}:
\begin{eqnarray}
&&\!\!\!\!\!\!\!\!\!\!\!\!
{\cal M}_{_{F}}(\q,\p)\!=\!g^2\!\int\!\frac{d^2\k_{1\perp}}{(2\pi)^2}
\frac{d^2\k_\perp}{(2\pi)^2}
\frac{\rho_{p,a}(\k_{1\perp})}{k_{1\perp}^2}
\!\int\!\! d^2\x_\perp d^2\y_\perp
e^{i\k_\perp\cdot\x_\perp}
e^{i(\p_\perp\!+\!\q_\perp\!-\!\k_\perp\!-\!\k_{1\perp})\cdot\y_\perp}
\nonumber\\
&&\times\Bigg\{
\frac
{\overline{u}(\q)\gamma^+(\slq-\slk+m)\gamma^-
(\slq-\slk-\slk_1+m)\gamma^+
[{\wt U}(\x_\perp)t^a {\wt U}^\dagger(\y_\perp)]
v(\p)}
{2p^+[(\q_\perp-\k_\perp)^2+m^2]+2q^+[(\q_\perp-\k_\perp-\k_{1\perp})^2+m^2]}
\nonumber\\
&&\quad+
\overline{u}(\q)\left[
\frac{\slC_{_{U}}(p+q,\k_{1\perp})}{(p+q)^2}
-
\frac{\gamma^+}{p^++q^+}
\right]t^b v(\p)\, U^{ba}(\x_\perp)
\Bigg\}\; .
\label{eq:Mf-final}
\end{eqnarray}
This is our final formula for the time ordered pair production
amplitude. 

\subsection{Physical interpretation and limits of  eq.~(\ref{eq:Mf-final})}
The interpretation of the two terms in this amplitude can
be made more transparent if we note that 
\begin{equation}
\overline{u}(\q)\left[
\frac{\slC_{_{U}}(p+q,\k_{1\perp})}{(p+q)^2}
-
\frac{\gamma^+}{p^++q^+}
\right]t^b v(\p)
=\overline{u}(\q)
\frac{\slC_{_{L}}(p+q,\k_{1\perp})}{(p+q)^2}
t^b v(\p)\; ,
\end{equation}
where $C_{_{L}}^\mu(p+q,\k_{1\perp})$ is the effective Lipatov vertex
for the production of a gluon of momentum $p+q$ via the fusion of two
gluons of momenta $k_1$ and $p+q-k_1$. 

We can then interpret the first term of eq.~(\ref{eq:Mf-final}) as the
production of a $q\bar{q}$ pair by a gluon from the proton {\sl
  before} it goes through the nucleus. It is therefore the $q\bar{q}$
state that interacts with the nucleus, and hence the two Wilson lines
in the fundamental representation at two different transverse
coordinates. The second term in eq.~(\ref{eq:Mf-final}) corresponds to
a gluon from the proton going through the nucleus before producing the
$q\bar{q}$ pair.  We recognize the Lipatov vertex that is known to
appear in the gluon production amplitude in pA collisions (see the
discussion in I) and the adjoint Wilson line $U$ describes the
rescatterings of this gluon while it goes through the nucleus before
it finally produces the pair outside the nucleus.

It is straightforward to verify that eq.~(\ref{eq:Mf-final}) agrees
with the expression that we obtained in \cite{GelisV1} in the limit of
low nuclear density (see eqs.~(15), (19) and (20) of \cite{GelisV1}),
namely, when we expand all the Wilson lines of eq.~(\ref{eq:Mf-final})
to first order in the source $\rho_{_{A}}$.

The amplitude in eq.~(\ref{eq:Mf-final}) satisfies another important
property: there is a cancellation between the two terms of
eq.~(\ref{eq:Mf-final}) in the limit $k_{1\perp}\to 0$. Indeed, for
$k_{1\perp}=0$, the curly bracket of eq.~(\ref{eq:Mf-final}) becomes
\begin{equation}
\big\{\cdots\big\}
=
\frac{\overline{u}(\q)\gamma^+
[{\wt U}(\x_\perp)t^a{\wt U}^\dagger(\y_\perp)]v(\p)}{p^++q^+}
-
\frac{\overline{u}(\q)\gamma^+[t^b U^{ba}(\x_\perp)]v(\p)}{p^++q^+}\; .
\end{equation}
Since this expression does not depend on $\k_\perp$ any more, the
integration over $\k_\perp$ in eq.~(\ref{eq:Mf-final}) produces a
$\delta(\x_\perp-\y_\perp)$. We can therefore replace
${\wt U}(\x_\perp)t^a{\wt U}^\dagger(\y_\perp)$ in the first term by
$t^bU^{ba}(\x_\perp)$ (thanks to eq.~(\ref{eq:wilson-alg-1}))
-- which proves the announced cancellation. {\it This property ensures
  collinear factorizability on the proton side\footnote{This property
    has been assumed and exploited in several recent studies of pA
    collisions~\cite{DumitJ1,DumitJ2,GelisJ1,GelisJ2,GelisJ3}.}, by
  making the $\k_{1\perp}$ integral at most logarithmically singular.}

\section{Pair production cross-section}
\label{eq:cs}
\subsection{Average over the color sources}
Before we square the amplitude in order to obtain the pair production
cross-section, it is convenient to write it in a more compact form by
introducing some shorthand notation. We shall write the amplitude as
\begin{eqnarray}
&&\!\!\!\!\!\!\!\!\!\!\!\!
{\cal M}_{_{F}}(\q,\p)\!=\!g^2\!\int\!\frac{d^2\k_{1\perp}}{(2\pi)^2}
\frac{d^2\k_\perp}{(2\pi)^2}
\frac{\rho_{p,a}(\k_{1\perp})}{k_{1\perp}^2}
\!\int\!\! d^2\x_\perp d^2\y_\perp
e^{i\k_\perp\cdot\x_\perp}
e^{i(\p_\perp\!+\!\q_\perp\!-\!\k_\perp\!-\!\k_{1\perp})\cdot\y_\perp}
\nonumber\\
&&\!\!\!\!\!\!\!\!\!\!\!\times
\overline{u}(\q)\Bigg\{ T_{q\bar{q}}(\k_{1\perp},\k_{\perp})
[{\wt U}(\x_\perp)t^a {\wt U}^\dagger(\y_\perp)]
+T_{g}(\k_{1\perp})[t^bU^{ba}(\x_\perp)]\Bigg\} v(\p)
\; ,\nonumber\\
&&
\label{eq:Mf-final-1}
\end{eqnarray}
where we denote\footnote{The momenta $p$ and $q$ of the produced
  particles have not been listed among the arguments of these objects
  in order to make the equations more compact.}
\begin{eqnarray}
&&T_{q\bar{q}}(\k_{1\perp},\k_{\perp})\equiv 
\frac{\gamma^+(\slq-\slk+m)\gamma^-(\slq-\slk-\slk_1+m)\gamma^+}
{2p^+[(\q_\perp\!-\!\k_\perp)^2+m^2]+2q^+[(\q_\perp\!-\!\k_\perp\!-\!\k_{1\perp})^2+m^2]}
\nonumber\\
&&T_{g}(\k_{1\perp})\equiv 
\frac{\slC_{_{L}}(p+q,\k_{1\perp})}{(p+q)^2}
\; .
\label{eq:Tqqbar-Tg}
\end{eqnarray}
Squaring the amplitude, and averaging over the configurations of the
color sources $\rho_p$ and $\rho_{_{A}}$, we have the following expression
for the differential probability to produce a pair in a collision at
impact parameter $\b$:
\begin{eqnarray}
&&\frac{dP_1(\b)}{d^2\p_\perp d^2\q_\perp dy_p dy_q}=\frac{g^4}{(16\pi^3)^2}
\int\limits_{\k_{1\perp},\k_{1\perp}^\prime,\k_{\perp},\k_{\perp}^\prime}
\frac{\left<\rho_{p,a}(\k_{1\perp})\rho_{p,a^\prime}^\dagger(\k_{1\perp}^\prime)\right>}
{k_{1\perp}^2 k_{1\perp}^{\prime 2}}
\nonumber\\
&&\!\!\times
\int\limits_{\x_\perp,\x_\perp^\prime,\y_\perp,\y_\perp^\prime}
\!\!\!\!\!
e^{i(\k_\perp\cdot\x_\perp-\k_\perp^\prime\cdot\x_\perp^\prime)}
e^{i(\p_\perp+\q_\perp-\k_\perp-\k_{1\perp})\cdot\y_\perp}
e^{-i(\p_\perp+\q_\perp-\k_\perp^\prime-\k_{1\perp}^\prime)\cdot\y_\perp^\prime}\nonumber\\
&&\!\!\times\Big\{
{\rm tr}_{\rm d}
\Big[(\slq\!+\!m)T_{q\bar{q}}(\slp\!-\!m)\gamma^0 T_{q\bar{q}}^{\dagger\prime}\gamma^0\Big]
{\rm tr}_{\rm c}\left<{\wt U}(\x_\perp)t^a {\wt U}^\dagger(\y_\perp)
{\wt U}(\y_\perp^\prime)t^{a^\prime}{\wt U}^\dagger(\x_\perp^\prime)
\right>
\nonumber\\
&&\!\!\;\;\,
+{\rm tr}_{\rm d}
\Big[(\slq\!+\!m)T_{q\bar{q}}(\slp\!-\!m)\gamma^0 T_{g}^{\dagger\prime}\gamma^0\Big]
{\rm tr}_{\rm c}\left<{\wt U}(\x_\perp)t^a {\wt U}^\dagger(\y_\perp)
t^{b^\prime}U^{\dagger{a^\prime b^\prime}}(\x_\perp^\prime)
\right>\nonumber\\
&&\!\!\;\;\,
+{\rm tr}_{\rm d}
\Big[(\slq\!+\!m)T_{g}(\slp\!-\!m)\gamma^0 T_{q\bar{q}}^{\dagger\prime}\gamma^0\Big]
{\rm tr}_{\rm c}\left<t^b U^{ba}(\x_\perp)
{\wt U}(\y_\perp^\prime)t^{a^\prime}{\wt U}^\dagger(\x_\perp^\prime)
\right>\nonumber\\
&&\!\!\;\;\,
+{\rm tr}_{\rm d}
\Big[(\slq\!+\!m)T_{g}(\slp\!-\!m)\gamma^0 T_{g}^{\dagger\prime}\gamma^0\Big]
{\rm tr}_{\rm c}\left<t^b U^{ba}(\x_\perp)
t^{b^\prime}U^{\dagger{a^\prime b^\prime}}(\x_\perp^\prime)
\right>
\Big\}\; .
\label{eq:P1-tmp-1}
\end{eqnarray}
In this formula, ${\rm tr}_{\rm d}$ denotes a trace over the Dirac
indices (we sum over all the spin states of the produced fermions),
${\rm tr}_{\rm c}$ is a trace over the color indices (we sum over the
color of the produced quark and antiquark). We have omitted the
arguments in $T_{q\bar{q}}$ and $T_g$, and $T_{q\bar{q}}^\prime$,
$T_g^\prime$ denote the same quantities with $\k_{1\perp},\k_\perp$
replaced by $\k_{1\perp}^\prime,\k_\perp^\prime$. We can already see
that, in general, this probability depends on 2-, 3- and 4-point
correlators of Wilson lines. As in I, we can introduce
the proton unintegrated gluon distribution as follows:
\begin{eqnarray}
&&g^2\left<\rho_{p,a}(\k_{1\perp})
\rho_{p,a^\prime}^\dagger(\k_{1\perp}^\prime)\right>=
\nonumber\\
&&\qquad=\frac{\delta^{aa^\prime}}{\pi d_{_{A}}}
\left[\frac{\k_{1\perp}+\k_{1\perp}^\prime}{2}\right]^2
\!\!\int\limits_{\X_\perp}\!\!
e^{i(\k_{1\perp}-\k_{1\perp}^\prime)\cdot \X_\perp}
\frac{d\varphi_p(\frac{\k_{1\perp}+\k_{1\perp}^\prime}{2}|\X_\perp)}
{d^2\X_\perp}\; .
\end{eqnarray}
The integration over $\X_\perp$ runs over the transverse profile of
the proton. $d_{_{A}}\equiv N^2-1$ is the dimension of the adjoint
representation of $SU(N)$, and $d\varphi_p/d^2\X_\perp$ is the
number of gluons per unit area and unit of transverse momentum in the
proton (namely, the proton non-integrated gluon distribution per unit
area). As long as we are interested in transverse momenta which are
large compared to the inverse proton size, we can approximate
$\k_{1\perp}\approx\k_{1\perp}^\prime$ and write
\begin{equation}
g^2\left<\rho_{p,a}(\k_{1\perp})
\rho_{p,a^\prime}^\dagger(\k_{1\perp}^\prime)\right>=
\frac{\delta^{aa^\prime}}{\pi d_{_{A}}}\k_{1\perp}^2
\!\!\int\limits_{\X_\perp}\!\!
e^{i(\k_{1\perp}-\k_{1\perp}^\prime)\cdot \X_\perp}
\frac{d\varphi_p(\k_{1\perp}|\X_\perp)}{d^2\X_\perp}\; .
\label{eq:phi-proton}
\end{equation}
We will also neglect the difference $\k_{1\perp}-\k_{1\perp}^\prime$
in the other factors of eq.~(\ref{eq:P1-tmp-1}). 

Let us now turn to the various correlators of Wilson lines that appear in
eq.~(\ref{eq:P1-tmp-1}). In eq.~(\ref{eq:P1-tmp-1}), the coordinates
$\x_\perp,\x_\perp^\prime,\y_\perp,\y_\perp^\prime$ are relative to
the center of the proton. Shifting all of them by the impact parameter
$\b$ makes them relative to the center of the nucleus, at the expense
of introducing an extra factor
$\exp(i(\k_{1\perp}-\k_{1\perp}^\prime)\cdot\b)$ in
eq.~(\ref{eq:P1-tmp-1}).  A generic property of these correlators of
Wilson lines is that, for a large nucleus, they are approximately
invariant by translation in the transverse plane: the violations of
translation invariance arise when the separation between two
coordinates become comparable to the nuclear radius. 

The simplest term in eq.~(\ref{eq:P1-tmp-1}) is the fourth one,
involving a correlator between two Wilson lines in the adjoint
representation. By analogy with eq.~(\ref{eq:phi-proton}), we can
define an unintegrated gluon distribution
$d\varphi_{_A}^{g,g}/d^2\X_\perp$ for the nucleus as
follows:
\begin{eqnarray}
&&\delta^{aa^\prime}
\int\limits_{\x_\perp,\x_\perp^\prime}
e^{i(\l_\perp\cdot\x_\perp-\l_\perp^\prime\cdot\x_\perp^\prime)}
{\rm tr}_{\rm c}\left<t^b U^{ba}(\x_\perp)
t^{b^\prime}U^{\dagger{a^\prime b^\prime}}(\x_\perp^\prime)
\right>
\nonumber\\
&&\qquad=
\frac{1}{2}\int\limits_{\x_\perp,\x_\perp^\prime}
e^{i(\l_\perp\cdot\x_\perp-\l_\perp^\prime\cdot\x_\perp^\prime)}
\left<U(\x_\perp)U^\dagger(\x_\perp^\prime)
\right>_{bb}\nonumber\\
&&\qquad=
\frac{g^2 N}{2\pi \l_\perp^2}
\int\limits_{\X_\perp}e^{i(\l_\perp-\l_\perp^\prime)\cdot\X_\perp}
\frac{d\varphi_{_A}^{g,g}(\l_\perp|\X_\perp)}{d^2\X_\perp}\; .
\label{eq:phi-gg}
\end{eqnarray}
Again, we have neglected $\l_\perp-\l_\perp^\prime$ in these
manipulations, because this difference is of the order of the inverse
of the radius of the nucleus or smaller. This equation can be seen as
a definition of the unintegrated gluon distribution in the nucleus,
that agrees at leading order in the source $\rho_{_{A}}$ with what we
have used for the proton. It is a natural object to use in this
problem because it absorbs all the rescattering effects of the gluon
on the nucleus. In fact, the same quantity already appeared for gluon
production in pA collisions\footnote{$\varphi_{_A}^{g,g}$ is identical
to the $\varphi_{_A}$ introduced in section 4.1 of I. The superscript
$g,g$ has been introduced for reasons that will become obvious
later.}.  Note that this is not the expectation value of the canonical
number operator (except at leading order in the source
$\rho_{_{A}}$). Up to some trivial factors, this object is the square
of the scattering amplitude of a gluon on the nucleus.

In order to deal with the 3-point correlator, it is convenient to define
a function $d\varphi_{_A}^{q\bar{q},g}/d^2\X_\perp$ such that:
\begin{eqnarray}
&&
\delta^{aa^\prime}
\!\!\!\!\!
\int\limits_{\x_\perp,\y_\perp,\x_\perp^\prime}
\!\!\!\!\!
e^{i(\k_\perp\cdot\x_\perp-\l_\perp^\prime\cdot\x_\perp^\prime)}
e^{i(\l_\perp-\k_\perp)\cdot\y_\perp}
{\rm tr}_{\rm c}\left<{\wt U}(\x_\perp)t^a {\wt U}^\dagger(\y_\perp)
t^{b^\prime}U^{\dagger{a^\prime b^\prime}}(\x_\perp^\prime)
\right>
\nonumber\\
&&\qquad\equiv
\frac{g^2 N}{2\pi \l_\perp^2}
\int\limits_{\X_\perp}e^{i(\l_\perp-\l_\perp^\prime)\cdot\X_\perp}
\frac{d\varphi_{_A}^{q\bar{q},g}(\k_\perp,\l_\perp-\k_\perp;\l_\perp|\X_\perp)}{d^2\X_\perp}\; .
\end{eqnarray}
Note that by construction, we have
\begin{equation}
\int\limits_{\k_\perp}
\frac{d\varphi_{_A}^{q\bar{q},g}(\k_\perp,\l_\perp-\k_\perp;\l_\perp|\X_\perp)}{d^2\X_\perp}
=\frac{d\varphi_{_A}^{g,g}(\l_\perp|\X_\perp)}{d^2\X_\perp}\; .
\label{eq:int-1}
\end{equation}
Indeed, integrating over $\k_\perp$ forces the coordinates $\x_\perp$
and $\y_\perp$ of the components of the $q\bar{q}$ pair to become
equal. When this occurs, the $q\bar{q}$ pair in an octet state is
effectively indistinguishable from a gluon. Mathematically, this
statement is a consequence of the identity in  eq.~(\ref{eq:wilson-alg-1}).
One can similarly write the 4-point correlator as follows
\begin{eqnarray}
&&
\delta^{aa^\prime}
\!\!\!\!\!
\int\limits_{\x_\perp,\y_\perp,\x_\perp^\prime,\y_\perp^\prime}
\!\!\!\!\!
e^{i(\k_\perp\cdot\x_\perp-\k_\perp^\prime\cdot\x_\perp^\prime)}
e^{i(\l_\perp-\k_\perp)\cdot\y_\perp}
e^{-i(\l_\perp^\prime-\k_\perp^\prime)\cdot\y_\perp^\prime}
\nonumber\\
&&\qquad\qquad\qquad\qquad\smash{\times
{\rm tr}_{\rm c}\left<{\wt U}(\x_\perp)t^a {\wt U}^\dagger(\y_\perp)
{\wt U}(\y_\perp^\prime)t^{a^\prime}{\wt U}^\dagger(\x_\perp^\prime)
\right>}
\nonumber\\
&&\nonumber\\
&&\equiv
\frac{g^2 N}{2\pi \l_\perp^2}
\int\limits_{\X_\perp}e^{i(\l_\perp-\l_\perp^\prime)\cdot\X_\perp}
\frac{d\varphi_{_A}^{q\bar{q},q\bar{q}}(\k_\perp,\l_\perp-\k_\perp;\k_\perp^\prime,\l_\perp-\k_\perp^\prime|\X_\perp)}{d^2\X_\perp}\; .
\end{eqnarray}
Note further that this definition is such that
\begin{eqnarray}
&&\int\limits_{\k_\perp,\k_\perp^\prime}
\frac{d\varphi_{_A}^{q\bar{q},q\bar{q}}(\k_\perp,\l_\perp-\k_\perp;\k_\perp^\prime,\l_\perp-\k_\perp^\prime|\X_\perp)}{d^2\X_\perp}
\nonumber\\
&&\quad
=
\int\limits_{\k_\perp}
\frac{d\varphi_{_A}^{q\bar{q},g}(\k_\perp,\l_\perp-\k_\perp;\l_\perp|\X_\perp)}{d^2\X_\perp}
=
\frac{d\varphi_{_A}^{g,g}(\l_\perp|\X_\perp)}{d^2\X_\perp}\; .
\label{eq:int-2}
\end{eqnarray}
In other words, collapsing one $q\bar{q}$ pair to a single point
restores the 3-point correlator, and collapsing each pair to a point
restores the 2-point correlator.

In terms of the functions we have just introduced, we can write the
differential production probability as follows:
\begin{eqnarray}
&&\frac{dP_1(\b)}{d^2\p_\perp d^2\q_\perp dy_p dy_q}=
\frac{\alpha_s^2 N}{8\pi^4 d_{_{A}}}
\int\limits_{\k_{1\perp},\k_{1\perp}^\prime,\k_{2\perp}}
\delta(\p_\perp+\q_\perp-\k_{1\perp}-\k_{2\perp})
\nonumber\\
&&\!\!\!\!\!\!\!\!\times
\int\limits_{\X_\perp,\Y_\perp}
\frac{e^{i(\k_{1\perp}-\k_{1\perp}^\prime)\cdot(\X_\perp-\Y_\perp+\b)}}{\k_{1\perp}^2 \k_{2\perp}^2}\;
\frac{d\varphi_p(\k_{1\perp}|\X_\perp)}{d^2\X_\perp}
\nonumber\\
&&\!\!\!\!\!\!\!\!
\times\Bigg\{
\int\limits_{\k_\perp,\k_\perp^\prime}
\!\!\!\!\!
{\rm tr}_{\rm d}
\Big[(\slq\!+\!m)T_{q\bar{q}}(\slp\!-\!m)
\gamma^0 T_{q\bar{q}}^{\dagger\prime}\gamma^0\Big]
\frac{d\varphi_{_A}^{q\bar{q},q\bar{q}}
(\k_\perp,\k_{2\perp}\!\!-\!\k_\perp;
\k_\perp^\prime,\k_{2\perp}\!\!-\!\k_\perp^\prime|\Y_\perp)}{d^2\X_\perp}
\nonumber\\
&&\!\!\!\;\;
+\int\limits_{\k_\perp}
\!
{\rm tr}_{\rm d}
\Big[(\slq\!+\!m)T_{q\bar{q}}(\slp\!-\!m)
\gamma^0 T_{g}^{\dagger\prime}\gamma^0\Big]
\frac{d\varphi_{_A}^{q\bar{q},g}
(\k_\perp,\k_{2\perp}\!-\!\k_\perp;\k_{2\perp}|\Y_\perp)}{d^2\X_\perp}
\nonumber\\
&&\!\!\!\;\;
+\int\limits_{\k_\perp^\prime}
\!{\rm tr}_{\rm d}
\Big[(\slq\!+\!m)T_{g}(\slp\!-\!m)\gamma^0 T_{q\bar{q}}^{\dagger\prime}\gamma^0\Big]
\frac{d\varphi_{_A}^{q\bar{q},g}
(\k_\perp^\prime,\k_{2\perp}\!-\!\k_\perp^\prime;\k_{2\perp}|\Y_\perp)}{d^2\X_\perp}
\nonumber\\
&&\!\!\!\qquad
+{\rm tr}_{\rm d}
\Big[(\slq\!+\!m)T_{g}(\slp\!-\!m)\gamma^0 T_{g}^{\dagger\prime}\gamma^0\Big]
\frac{d\varphi_{_A}^{g,g}(\k_{2\perp}|\Y_\perp)}{d^2\X_\perp}
\Bigg\}\; .
\label{eq:P1-tmp-2}
\end{eqnarray}
At this point, it is straightforward to integrate over the impact
parameter in order to turn the probability $P_1(\b)$ into a
cross-section. This integration produces a
$\delta(\k_{1\perp}-\k_{1\perp}^\prime)$, which makes the integration
over $\k_{1\perp}^\prime$ trivial. Then, integrating over the
coordinates $\X_\perp$ and $\Y_\perp$ can be done independently in the
proton and in the nucleus, which restores the $\varphi_p$ and the
$\varphi_{_A}$ functions for the whole projectiles rather than per
unit area. Our final expression for the differential cross-section of
quark pairs produced in proton-nucleus collisions is
\begin{eqnarray}
&&\frac{d\sigma}{d^2\p_\perp d^2\q_\perp dy_p dy_q}=
\frac{\alpha_s^2 N}{8\pi^4 d_{_{A}}}
\int\limits_{\k_{1\perp},\k_{2\perp}}
\frac{\delta(\p_\perp+\q_\perp-\k_{1\perp}-\k_{2\perp})}
{\k_{1\perp}^2 \k_{2\perp}^2}
\nonumber\\
&&\!\!\!\!\!
\times\Bigg\{
\int\limits_{\k_\perp,\k_\perp^\prime}
\!\!\!\!\!
{\rm tr}_{\rm d}
\Big[(\slq\!+\!m)T_{q\bar{q}}(\slp\!-\!m)
\gamma^0 T_{q\bar{q}}^{\prime\dagger}\gamma^0\Big]
\varphi_{_A}^{q\bar{q},q\bar{q}}
(\k_\perp,\k_{2\perp}\!-\!\k_\perp;
\k_\perp^\prime,\k_{2\perp}\!-\!\k_\perp^\prime)
\nonumber\\
&&\;\;
+\int\limits_{\k_\perp}
\!
{\rm tr}_{\rm d}
\Big[(\slq\!+\!m)T_{q\bar{q}}(\slp\!-\!m)
\gamma^0 T_{g}^{\dagger}\gamma^0\Big]
\varphi_{_A}^{q\bar{q},g}
(\k_\perp,\k_{2\perp}\!-\!\k_\perp;\k_{2\perp})
\nonumber\\
&&\;\;
+\int\limits_{\k_\perp}
\!{\rm tr}_{\rm d}
\Big[(\slq\!+\!m)T_{g}(\slp\!-\!m)\gamma^0 T_{q\bar{q}}^{\dagger}\gamma^0\Big]
\varphi_{_A}^{q\bar{q},g}
(\k_\perp,\k_{2\perp}\!-\!\k_\perp;\k_{2\perp})
\nonumber\\
&&\qquad
+{\rm tr}_{\rm d}
\Big[(\slq\!+\!m)T_{g}(\slp\!-\!m)\gamma^0 T_{g}^{\dagger}\gamma^0\Big]
\varphi_{_A}^{g,g}(\k_{2\perp})
\Bigg\}
\varphi_p(\k_{1\perp})\; .
\label{eq:cross-section}
\end{eqnarray}
This expression for pair production is the main result of this paper.  Note that since  $\k_{1\perp}=\k_{1\perp}^\prime$, the
abbreviations $T_{q\bar{q}}, T_{q\bar{q}}^\prime, T_g$ are related to
eqs.~(\ref{eq:Tqqbar-Tg}) as follows:
\begin{eqnarray}
&& T_{q\bar{q}}\equiv T_{q\bar{q}}(\k_{1\perp},\k_{\perp})\;,\quad
T_{q\bar{q}}^\prime
\equiv T_{q\bar{q}}(\k_{1\perp},\k_{\perp}^\prime)\; ,\nonumber
\nonumber\\
&&T_g\equiv T_{g}(\k_{1\perp})\; .
\end{eqnarray}
The Dirac traces in eq.~\ref{eq:cross-section} can be computed quite
straightforwardly using a symbolic manipulation program like {\sc
FORM}.  We will give explicit expressions for these in a future paper
where we explore the phenomenological consequences of our result.

\subsection{Leading twist approximation}
The unintegrated distributions that we encountered in the previous
section contain contributions from all twists and are in general quite
complicated.  As discussed briefly in the introduction, they can be
obtained from numerical solutions of the JIMWLK
equations~\cite{RummuW1}. For Gaussian correlations, these can be
computed analytically, and an explicit derivation is given in appendix
\ref{sec:4-point-corr}.  Here, we will discuss only the leading twist
approximation, which has a simple physical interpretation and is
useful in order to compare eq.~(\ref{eq:cross-section}) with the
result obtained for $pp$ collisions.

At leading order in the density $\mu_{_{A}}^2$ that describes the
nucleus, it is easy to derive a closed form expression for the functions
$\varphi_{_A}^{g,g}$, $\varphi_{_A}^{g,q\bar{q}}$ and
$\varphi_{_A}^{q\bar{q},q\bar{q}}$ introduced in the previous section.
The only ingredient needed for this leading order calculation is the
average of two $\rho_{_{A}}$'s, namely,
\begin{equation}
\left<\rho_{_{A},a}(\x_\perp)\rho_{_A,a^\prime}(\x_\perp^\prime)\right>
=
\delta_{aa^\prime}\delta(\x_\perp-\x_\perp^\prime)\mu_{_{A}}^2\; .
\end{equation}

From the definitions given in the previous section, one then obtains
the following relations:
\begin{eqnarray}
&&\varphi_{_A}^{g,g}(\k_{2\perp})\approx
\pi R^2 \left[\pi d_{_{A}} g^2 \frac{\mu_{_{A}}^2}{\k_{2\perp}^2}\right]\; ,
\nonumber\\
&&
\varphi_{_A}^{g,q\bar{q}}(\k_\perp,\k_{2\perp}-\k_\perp|\k_{2\perp})
\approx
\frac{\pi R^2}{2}
 \left[\pi d_{_{A}} g^2 \frac{\mu_{_{A}}^2}{\k_{2\perp}^2}\right]
(2\pi)^2\left[\delta(\k_\perp)\!+\!\delta(\k_\perp\!-\!\k_{2\perp})\right]
\nonumber\\
&&
\varphi_{_A}^{q\bar{q},q\bar{q}}(\k_\perp,\k_{2\perp}-\k_\perp|\k_\perp^\prime,\k_{2\perp}-\k_\perp^\prime)\approx
\pi R^2 \left[\pi d_{_{A}} g^2 \frac{\mu_{_{A}}^2}{\k_{2\perp}^2}\right]
\nonumber\\
&&\qquad\qquad\qquad\qquad\times
(2\pi)^4\Big[
\frac{C_{_{F}}}{N} \left(
\delta(\k_\perp-\k_{2\perp})\delta(\k_\perp^\prime-\k_{2\perp})
+
\delta(\k_\perp)\delta(\k_\perp^\prime)
\right)
\nonumber\\
&&\qquad\qquad\qquad\qquad\qquad
+\frac{1}{2N^2}
\left(
\delta(\k_\perp-\k_{2\perp})\delta(\k_\perp^\prime)
+
\delta(\k_\perp)\delta(\k_\perp^\prime-\k_{2\perp})
\right)
\Big]\; .\nonumber\\
&&
\end{eqnarray}
In the derivation of these relations, we have assumed that the nuclei
have a large transverse area $\pi R^2$ and that their density is
nearly uniform in the transverse plane.  If one substitutes these
leading twist approximations in eq.~(\ref{eq:cross-section}), it is
straightforward to re-obtain the expression of the cross-section we
derived in a previous paper for $pp$ collisions (see eqs.~(42), (B5),
(B6) and (B7) of ref.~\cite{GelisV1}).

Note also that the leading twist expressions for
$\varphi_{_{A}}^{g,q\bar{q}}$ and $\varphi_{_A}^{q\bar{q},q\bar{q}}$
have a simple interpretation. At this order, the probe (gluon or
$q\bar{q}$ pair) interacts with the nucleus by a single gluon
exchange. The two delta functions in $\varphi_{_{A}}^{g,q\bar{q}}$
correspond to the gluon being attached to the antiquark line
($\delta(\k_\perp)$ -- there is no momentum flow from the nucleus to
the quark line) or to the quark line ($\delta(\k_{2\perp}-\k_\perp)$
-- all the momentum from the nucleus flows on the quark line). An
identical interpretation holds for the four terms of
$\varphi_{_A}^{q\bar{q},q\bar{q}}$.

\subsection{$k_\perp$-factorization for pair production ?}
From eq.~(\ref{eq:cross-section}), one sees immediately that in
general we cannot write the pair production cross-section for pA
collisions in a $k_\perp$-factorized manner, at least not in the usual
sense. The closest to $k_\perp$-factorization we can achieve for pA
collisions is precisely eq.~(\ref{eq:cross-section}), where we see
that we need three different functions in order to describe the
nucleus. The physical reason for this is that a quark-antiquark pair
is a composite object: the amplitude for producing a pair depends on
how much momentum the nucleus gives to the quark and how much momentum it
gives to the antiquark (in case the pair is formed before going
through the nucleus), and not just on the total momentum taken to the
nucleus. This also explains why a true $k_\perp$-factorization is
possible when we treat the problem of pair production at leading order
in the source $\rho_{_{A}}$. Indeed, at this order, the pair can only
exchange one gluon with the nucleus, which is either attached to the
quark or to the antiquark, and a single momentum variable
 completely characterizes this exchange.

From this discussion, one can also foresee in what limit a true
$k_\perp$ factorization would be possible. From the integral relations
of eqs.~(\ref{eq:int-1}) and (\ref{eq:int-2}), we indeed see that if
we can neglect the dependence on $\k_\perp$ in
$T_{q\bar{q}}(\k_{1\perp},\k_\perp)$, then the integrations over
$\k_\perp$ and $\k_\perp^\prime$ act only on the functions
$\varphi_{_A}^{q\bar{q},q\bar{q}}$ and $\varphi_{_A}^{q\bar{q},g}$ and
simply give $\varphi_{_A}^{g,g}$ which we can now factor out.  Such an
approximation is valid when the momenta $\p_\perp,\q_\perp$ of the
produced quark and antiquark are large, or when the mass $m$ of the
produced quark is large. In this case, one effectively produces a
pair which has a small transverse size and behaves  like a
single gluon.

\section{Single quark cross-section}
\label{sec:single-quark}
We shall now compute the single quark inclusive
cross-section which is phenomenologically interesting in its own right. It is obtained from the pair production production
cross-section given in eq.~(\ref{eq:cross-section}) by integrating
over the transverse momentum $\p_\perp$ and rapidity $y_p$ of the
antiquark. Integrating over $\p_\perp$ is trivial thanks to the delta function 
$\delta(\p_\perp+\q_\perp-\k_{1\perp}-\k_{2\perp})$, and simply amounts
to replacing $\p_\perp\to\k_{1\perp}+\k_{2\perp}-\q_\perp$ in
the rest of the expression.

There is an interesting simplification of the first term, involving
the 4-point function $\varphi_{_A}^{q\bar{q},q\bar{q}}$. One notices
that $T_{q\bar{q}}$ (see the first of eqs.~(\ref{eq:Tqqbar-Tg})) does
not depend on $\p_\perp$.  Moreover, since $T_{q\bar{q}}$ has a
$\gamma^+$ on the right, the product $T_{q\bar{q}} (\slp-m)\gamma^0
T_{q\bar{q}}^{\prime\dagger}\gamma^0$ does not depend on $\p_\perp$
either (because $\gamma^+(\slp-m)\gamma^+=2p^+\gamma^+$ does not
depend on $\p_\perp$). Therefore, this object does not depend on
$\k_{2\perp}$ after the substitution
$\p_\perp\to\k_{1\perp}+\k_{2\perp}-\q_\perp$.  As a consequence, the
integration over $\k_{2\perp}$ can be performed directly for this
term, since we have
\begin{eqnarray}
&&\int\frac{d^2\k_{2\perp}}{(2\pi)^2}
\frac{1}{k_{2\perp}^2}
\varphi_{_A}^{q\bar{q},q\bar{q}}
(\k_\perp,\k_{2\perp}\!-\!\k_\perp;
\k_\perp^\prime,\k_{2\perp}\!-\!\k_\perp^\prime)
\nonumber\\
&&
\!\!\!\!\!\!\!\!\!
=\frac{2\pi C_{_{F}}}{g^2 N}
(2\pi)^2\delta(\k_\perp-\k_\perp^\prime)
\int\limits_{\x_\perp,\x_\perp^\prime}
e^{i\k_\perp\cdot(\x_\perp-\x_\perp^\prime)}
{\rm tr}_{\rm c}\left<
{\wt U}(\x_\perp){\wt U}^\dagger(\x_\perp^\prime)
\right>\; .
\end{eqnarray}
Thus we see that the 4-point function reduces to a much simpler 2-point
function.  It is then natural to define, by analogy with
eq.~(\ref{eq:phi-gg}), the following function:
\begin{equation}
\int\limits_{\x_\perp,\x_\perp^\prime}
e^{i\k_\perp\cdot(\x_\perp-\x_\perp^\prime)}
{\rm tr}_{\rm c}\left<
{\wt U}(\x_\perp){\wt U}^\dagger(\x_\perp^\prime)
\right>
\equiv 
\frac{g^2}{2\pi\k_\perp^2}
\varphi_{_A}^{q,q}(\k_\perp)\; .
\end{equation}
The normalization has been chosen such that this definition becomes
equivalent to eq.~(\ref{eq:phi-proton}) in the limit where the source
$\rho_{_{A}}$ which describes the nucleus is a weak source. Because
$T_g$ depends explicitly on $\p_\perp$, it becomes
$\k_{2\perp}$-dependent after the substitution
$\p_\perp\to\k_{1\perp}+\k_{2\perp}-\q_\perp$. For this reason, no
such simplification occurs for the last three terms of
eq.~(\ref{eq:cross-section}).  We can now write the single quark
production cross-section as follows:
\begin{eqnarray}
&&\frac{d\sigma_q}{d^2\q_\perp dy_q}=
\frac{\alpha_s^2 N}{8\pi^4 d_{_{A}}}\int\frac{dp^+}{p^+}
\int\limits_{\k_{1\perp},\k_{2\perp}}
\frac{1}{\k_{1\perp}^2 \k_{2\perp}^2}
\nonumber\\
&&\!\!\!\!\!
\times\Bigg\{
{\rm tr}_{\rm d}
\Big[(\slq\!+\!m)T_{q\bar{q}}(\k_{1\perp},\k_{2\perp})(\slp\!-\!m)
\gamma^0 T_{q\bar{q}}^{\dagger}(\k_{1\perp},\k_{2\perp})\gamma^0\Big]
\frac{C_{_{F}}}{N}\varphi_{_A}^{q,q}
(\k_{2\perp})
\nonumber\\
&&\!\!\!\!\!\!\!
+\int\limits_{\k_\perp}
\!
{\rm tr}_{\rm d}
\Big[(\slq\!+\!m)T_{q\bar{q}}(\k_{1\perp},\k_{\perp})(\slp\!-\!m)
\gamma^0 T_{g}^{\dagger}(\k_{1\perp})\gamma^0\Big]
\varphi_{_A}^{q\bar{q},g}
(\k_\perp,\k_{2\perp}\!-\!\k_\perp;\k_{2\perp})
\nonumber\\
&&\!\!\!\!\!\!\!
+\int\limits_{\k_\perp}
\!{\rm tr}_{\rm d}
\Big[(\slq\!+\!m)T_{g}(\k_{1\perp})(\slp\!-\!m)\gamma^0 T_{q\bar{q}}^{\dagger}(\k_{1\perp},\k_{\perp})\gamma^0\Big]
\varphi_{_A}^{q\bar{q},g}
(\k_\perp,\k_{2\perp}\!-\!\k_\perp;\k_{2\perp})
\nonumber\\
&&\;
+{\rm tr}_{\rm d}
\Big[(\slq\!+\!m)T_{g}(\k_{1\perp})(\slp\!-\!m)\gamma^0 T_{g}^{\dagger}(\k_{1\perp})\gamma^0\Big]
\varphi_{_A}^{g,g}(\k_{2\perp})
\Bigg\}
\varphi_p(\k_{1\perp})\; .
\label{eq:cross-section-q}
\end{eqnarray}
In the first term, we have re-labeled $\k_\perp$ as $\k_{2\perp}$ in
order to be able to factor out the term $1/\k_{2\perp}^2$.  Note also
that $\p_\perp$ should be understood as
$\k_{1\perp}+\k_{2\perp}-\q_\perp$ in this formula.  Obviously, we
cannot factor out in eq.~(\ref{eq:cross-section-q}) a common function
$\varphi_{_A}$ that would describe the gluon content of the
nucleus. This of course means that the single quark production
cross-section, even if simpler than the pair production cross-section,
is still not $k_\perp$-factorizable.

Recently, in ref.~\cite{Tuchi1}, Tuchin considered the inclusive
single quark distribution in proton-nucleus collisions for a model
where the collisions on the successive nucleons are independent. This
is equivalent to the case of Gaussian correlations in the MV-model.
Tuchin's discussion is in the light-cone gauge $A_+=0$, and is
expressed as the convolution of a) the square of the amplitude for the
emission of a gluon $g$ by a valence quark $q_v$ of the proton b) the
square of the amplitude for the splitting of this gluon into a
$q\bar{q}$ pair and c) the rescatterings of the $q_v g q \bar{q}$
system\footnote{In our expression, the valence quark of the proton
  that emitted the gluon does not appear in the correlators that
  describe the rescatterings on the nucleus. In our language, the
  rescatterings of this valence quark correspond to gauge rotations of
  the color current of the proton, and are taken care of by the
  Lipatov vertex.} on the nucleus. Since his results are expressed
entirely in coordinate space, Tuchin does not address the issue of
$k_\perp$-factorization. Further, since our results, in contrast, are
entirely formulated in momentum space, a detailed comparison of his
result with the Gaussian correlation limit of our general expression
is not at present feasible. Nevertheless, several features of his
result can be identified with terms appearing in our result.

\section{Conclusions}
We have computed in this paper the production of quark pairs in high
energy proton-nucleus collisions in the Color Glass Condensate
approach.  We find that the pair production cross-section, unlike the
inclusive gluon production cross-section discussed in I, is not
$k_\perp$-factorizable into a product of unintegrated distributions,
because the quark and the anti-quark can separately interact with the
nucleus after they are produced. The same is true also for the single
quark distribution. In the limit of very large mass pairs, or when the
quark and the anti-quark have large transverse momenta,
$k_\perp$-factorization is recovered.

Though not $k_\perp$-factorizable in the usual sense, the pair
production cross-section can still be expressed in terms of the
product of the unintegrated gluon distribution of the proton times the
sum of four terms containing four such unintegrated distributions
describing the nucleus. One of these four terms is the 2-point
correlator of two adjoint Wilson lines which already appeared in gluon
production in pA collisions. This distributions corresponds to a gluon
interacting with the nucleus in the amplitude and in the complex
conjugate amplitude. Two of the other terms contain a 3-point
correlator of two fundamental Wilson lines and an adjoint Wilson
line. This corresponds to a gluon in the amplitude and a
quark--anti-quark pair in the complex conjugate amplitude or vice
versa. The fourth unintegrated distribution is related to the 4-point
correlator of four fundamental Wilson lines; this of course
corresponds to quark--anti-quark pairs in both the amplitude and the
complex conjugate amplitude.

The single quark distribution is also not $k_\perp$-factorizable. A
significant simplification nevertheless occurs because the correlator
of 4 fundamental Wilson lines is replaced by the correlator of two
fundamental Wilson lines. One therefore has only contributions from
2-point and 3-point correlators of Wilson lines to the single quark
distribution.

The three and four point correlators that appear in pair production
cross-section probe non-trivial multi-parton correlations in addition
to those involved in inclusive gluon production alone.  The three
point function that appears here is the same as one that appears in
inelastic and diffractive deeply inelastic scattering at high
energies~\cite{Kovch6,KovneW1}. These correlators of Wilson lines are
therefore the relevant universal degrees of freedom at high energies,
appearing in all high energy scattering reactions.  The quantum
evolution of these N-point correlators of Wilson lines at small $x$ is
described by the JIMWLK equations.  Measurements of the pair
production and single quark distributions can therefore provide a
sensitive test of such evolution equations.

In the McLerran-Venugopalan model of Gaussian correlations, these 2,3
and 4-point correlators of Wilson lines can be computed
analytically. The explicit expressions for these are derived in
appendix B.  The 2-point function has been known for a long
time~\cite{JalilKMW1,KovchM3}. The 3-point function is, for Gaussian
correlations, the $q{\bar q}g$-propagator and was computed in
refs.~\cite{KovneW1,Kovch6,KopelST1,KopelTH1}. For single
quark production in the MV-model, as discussed here and by Tuchin
previously, it is only these correlators that appear in the
cross-section. For pair production, one needs the 4-point function as
well. This is computed for Gaussian correlations in appendix B.

 As we discussed extensively in I, the MV-model is a good model for
high-energy scattering when $x$ is not too small -- or quantum
evolution is not important. When quantum evolution sets in, the
initial condition of Gaussian correlations changes
quickly~\cite{AyalaJMV1,AyalaJMV2} with non-Gaussian correlations
becoming increasingly important at smaller values of $x$. The data at
RHIC suggest that the MV-model might provide a good description of
Deuteron-Gold collisions at central rapidities (when $x\sim 10^{-2}$)
but is a bad description already at forward rapidities corresponding
to $x\sim 10^{-3}$. Computations of single quark and pair-production
cross-section in both the MV-model and with the JIMWLK evolution
equations will greatly help confirm the picture of D-Au collisions
that is emerging from the RHIC data. We plan to pursue the
phenomenological implications of our results in a future publication.

\section*{Acknowledgments}
R.~V.'s research was supported by DOE Contract
No. DE-AC02-98CH10886. We would like to thank I.~Balitsky, E.~Iancu,
K.~Itakura, J.~Jalilian-Marian, K.~Kajantie, Yu.~Kovchegov, T.~Lappi,
S.~Munier, D.~Triantafyllopoulos, and K.~Tuchin for useful discussions
on this and related topics.

\appendix

\section{Gaussian averages of Wilson lines}
\label{sec:4-point-corr}
\subsection{Introduction}
In this appendix, we derive the expression of the correlators of
Wilson lines that appear in eq.~(\ref{eq:P1-tmp-1}). The most
complicated one is the 4-point function $C(\x_\perp,\y_\perp;\u_\perp,\v_\perp)
\equiv{\rm tr}_{\rm c}
\big<{\wt U}(\x_\perp)t^a {\wt U}^\dagger(\y_\perp)
{\wt U}(\u_\perp)t^a {\wt U}^\dagger(\v_\perp)\big>$, and it is
trivial to verify that the 3-point and 2-point correlators that also
appear in eq.~(\ref{eq:P1-tmp-1}) can be obtained as limits of the
4-point function:
\begin{eqnarray}
&&{\rm tr}_{\rm c}\left<{\wt U}(\x_\perp)t^a {\wt U}^\dagger(\y_\perp)
t^b U^{\dagger ab}(\u_\perp)
\right>=C(\x_\perp,\y_\perp;\u_\perp,\u_\perp)\nonumber\\
&&{\rm tr}_{\rm c}\left<
t^b U^{ba}(\x_\perp) t^{b^\prime}
U^{\dagger a b^\prime}(\u_\perp)
\right>
=
C(\x_\perp,\x_\perp;\u_\perp,\u_\perp)\; .
\end{eqnarray}
Thanks to these relations, it is sufficient to compute the 4-point
function. Similar correlators have been studied in \cite{KovneW1} and
\cite{Fujii1}, although the 4-point function we need here is not
derived explicitly in these papers. It is in general impossible to
obtain closed expressions for these correlators. An exception is the
case of a Gaussian weight function for the probability distribution of
the source $\rho_{_{A}}$:
\begin{equation}
W_{_{A}}[\rho_{_{A}}]\equiv
\exp\left[-
\int dx^+ d^2\x_\perp 
\frac{\overline\rho_{_{A},a}(x^+,\x_\perp)\overline\rho_{_{A},a}(x^+,\x_\perp)}
{2\mu_{_{A}}^2(x^+)}
\right]\; .
\label{eq:W}
\end{equation}
We denote\footnote{We must reintroduce the smearing in $x^+$ here in
order to make sense of the path ordering that appears in the Wilson
lines. It will however not play any role in the final results.}
$\overline{\rho}_2(x^+,\x_\perp)\equiv
\delta_\epsilon(x^+)\rho_{_{A}}(\x_\perp)$, and $\mu_{_{A}}^2(x^+)$ is
the density of color charges at a given $x^+$ (this function is
strongly peaked around $x^+=0$). The most elementary correlator is
\begin{equation}
\left<\rho_{_{A},a}(x^+,\x_\perp)\rho_{_{A},b}(y^+,\y_\perp)\right>
=
\delta_{ab}\delta(x^+-y^+)\delta(\x_\perp-\y_\perp)\mu_{_{A}}^2(x^+)\; ,
\end{equation}
which is local in $x^+$ and $\x_\perp$.  (The latter property is not
essential in the following calculation, and can be relaxed if necessary.)

In the calculation of the 4-point function
$C(\x_\perp,\y_\perp;\u_\perp,\v_\perp)$, the most complicated task is
to deal with the $SU(N)$ color algebra. A very convenient way to do
that is to use a semi-graphical method. We can represent pictorially
the correlator that we need as 
\setbox1=%
\hbox to 4cm{\hfil\resizebox*{4cm}{!}{\includegraphics{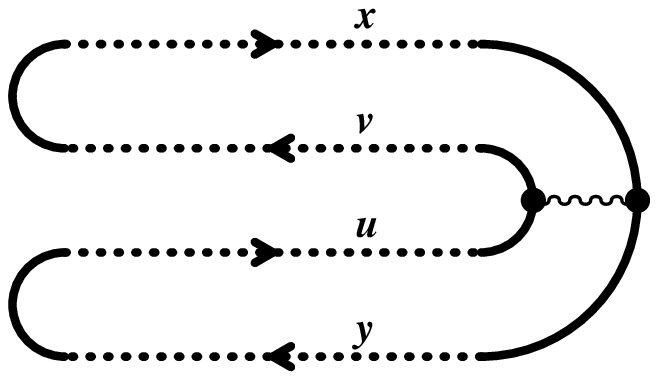}}}
\begin{equation}
C(\x_\perp,\y_\perp;\u_\perp,\v_\perp)=\quad\raise -9.5mm\box1\;\;\; .
\end{equation}
In this diagrammatic representation, the dotted lines represent Wilson
lines in the fundamental representation with the horizontal axis the
$x^+$ axis. Note that they therefore depend on the source
$\rho_{_{A}}$ to all orders.  The arrows indicate the direction of the
path-ordering, i.e. whether we have a ${\wt U}$ of a ${\wt
U}^\dagger$.  The thick black dots are color matrices $t^a$ in the
fundamental representation, and the wavy line that connects them
denotes a contraction of their color indices ($t^a\cdots t^a$). The
Wilson lines are connected at $x^+=\pm\infty$ by solid lines which
merely indicate how the Wilson lines are multiplied (these solid lines
do not contain any $\rho_{_{A}}$ nor color matrices). The closed loop
indicates that the product of Wilson lines is traced. An essential
ingredient in the forthcoming calculation is the Fierz identity in
$SU(N)$\footnote{See \cite{Cvita1} for a detailed discussion of group
algebra manipulations in non-abelian gauge theories.}:
\begin{equation}
t^a_{ij}t^a_{kl}
=
\frac{1}{2} \delta_{il}\delta_{jk}-\frac{1}{2N} \delta_{ij}\delta_{kl}\; ,
\end{equation}
which  has
the following graphical representation:
\setbox1=%
\hbox to 1.5cm{\hfil\resizebox*{!}{8mm}{\includegraphics{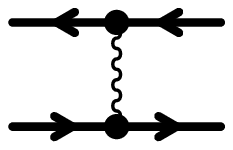}}}
\setbox2=%
\hbox to 1.5cm{\hfil\resizebox*{!}{8mm}{\includegraphics{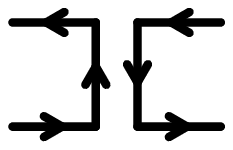}}}
\setbox3=%
\hbox to 0.8cm{\hfil\resizebox*{!}{8mm}{\includegraphics{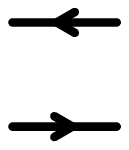}}}
\begin{equation}
\raise -3.5mm\box1
=\frac{1}{2}\raise -3.5mm\box2
-\frac{1}{2N}\raise -3.5mm\box3\;\;\; .
\label{eq:fierz}
\end{equation}
Thus, the 4-point correlator $C$ simplifies into a sum of two
simpler terms:
\setbox1=%
\hbox to 2.4cm{\hfil\resizebox*{2.4cm}{!}{\includegraphics{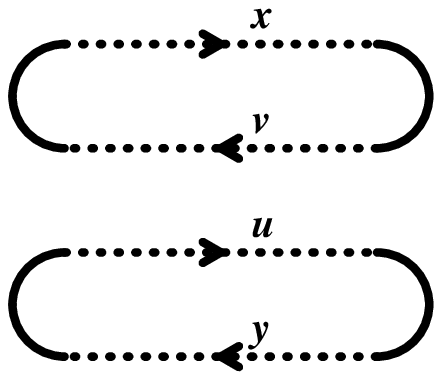}}}
\setbox2=%
\hbox to 3cm{\hfil\resizebox*{3cm}{!}{\includegraphics{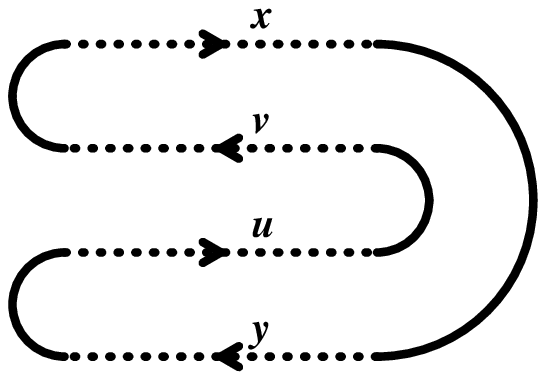}}}
\begin{equation}
C(\x_\perp,\y_\perp;\u_\perp,\v_\perp)=\;
\frac{1}{2}\;\raise -9mm\box1
\;-\frac{1}{2N}\;\raise -9mm\box2\;\; .
\label{eq:C-decomp} 
\end{equation}
It is therefore sufficient to calculate the following object of
$SU(N)\times SU(N)$:
\begin{equation}
{\cal M}_{ijkl}(\x_\perp,\y_\perp,\u_\perp,\v_\perp)\equiv
\left<
\left[{\wt U}^\dagger(\v_\perp){\wt U}(\x_\perp)\right]_{ij}
\left[{\wt U}^\dagger(\y_\perp){\wt U}(\u_\perp)\right]_{kl}
\right>\; ,
\end{equation}
which can be represented as a diagram with open ends:
\setbox1=%
\hbox to 2.4cm{\hfil\resizebox*{2.4cm}{!}{\includegraphics{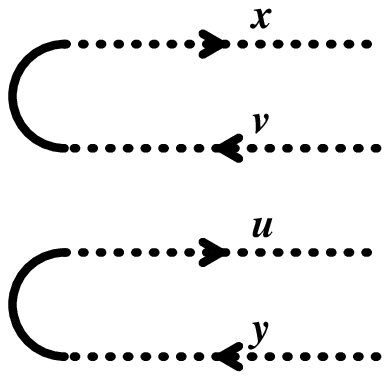}}}
\begin{equation}
{\cal M}_{ijkl}(\x_\perp,\y_\perp,\u_\perp,\v_\perp)
=\raise -10mm\box1\;\;\; .
\end{equation}
Once we know this object, the correlator $C$ is given by:
\begin{equation}
C(\x_\perp,\y_\perp;\u_\perp,\v_\perp)=
\frac{1}{2}
{\cal M}_{iikk}(\x_\perp,\y_\perp,\u_\perp,\v_\perp)
-
\frac{1}{2N}
{\cal M}_{ikki}(\x_\perp,\y_\perp,\u_\perp,\v_\perp)\; .
\end{equation}

\subsection{Tadpole corrections}

Thanks to the Gaussian form of the functional $W[\rho_{_{A}}]$, we can
expand ${\cal M}$ in powers of the elementary correlator
$\left<\rho_{_{A}}\rho_{_{A}}\right>$ using Wick's theorem. This
amounts to expanding the Wilson lines in powers of the source
$\rho_{_{A}}$ and to connecting the $\rho_{_{A}}$'s pairwise. From the
point of view of its color structure, such a pairing
$\left<\rho_{_{A}}\rho_{_{A}}\right>$ contains a contracted pair
$t^a\cdots t^a$ of color matrices and can be represented by two black
dots connected by a wavy link. Since
$\left<\rho_{_{A}}\rho_{_{A}}\right>$ is local in $x^+$, only links
that have their endpoints at the same $x^+$ are allowed.  Some of them
are ``tadpoles'' that start and end on the same Wilson line, and can
be represented as follows:
\setbox1=%
\hbox to 2.4cm{\hfil\resizebox*{2.4cm}{!}{\includegraphics{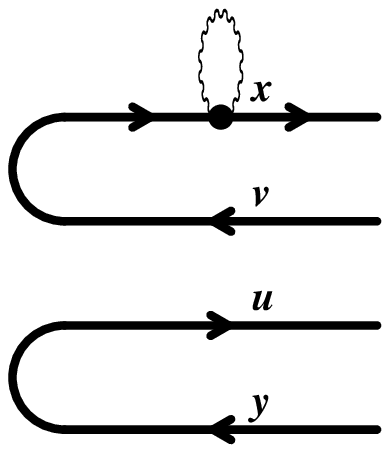}}}
\begin{equation}
\raise -10mm\box1\;\;\; .
\end{equation}
In this diagrammatic representation, the black dots are color
matrices\footnote{For tadpoles, there is one $t^a$ at each end of the
wavy line, even if only one dot appears in the diagram.}  $t^a$,
and the solid line simply indicates in what order they are multiplied.
A tadpole insertion on the Wilson line of transverse coordinate
$\x_\perp$ at the ``time'' $z^+$ brings a factor
$-C_{_{F}}\mu_{_{A}}^2(z^+) L(x,x)$, with:
\begin{equation}
L(x,x)\equiv g^4\int_{\z_\perp}G_0(\x_\perp-\z_\perp)^2\; ,
\label{eq:L-1}
\end{equation}
where $C_{_{F}}\equiv (N^2-1)/2N$ is the quadratic Casimir in the
fundamental representation of $SU(N)$ and where
\begin{equation}
G_0(\x_\perp-\z_\perp)\equiv \int\frac{d^2\k_\perp}{(2\pi)^2}
\frac{e^{i\k_\perp\cdot(\x_\perp-\z_\perp)}}{\k_\perp^2}
\end{equation}
is the 2-dimensional free massless propagator. Note that tadpole
insertions are color singlet, and therefore commute with the rest of
the expression. As a consequence, we can easily factor them out and
resum them. This gives a factor\footnote{The factor $1/2$ in the
exponential is due to the fact that the two ends of a link are ordered
in $z^+$ if they belong to the same Wilson line.}:
\begin{equation}
{\cal T}
\equiv e^{-\frac{1}{2}C_{_{F}}{\mu_{_{A}}^2}
\big[
{L}(x,x)+{L}(y,y)+{L}(u,u)+{L}(v,v)
\big]}\; ,
\end{equation}
where we denote the density integrated over the longitudinal coordinate:
\begin{equation}
\mu_{_{A}}^2\equiv\int\limits_{-\infty}^{+\infty}dz^+ \, \mu_{_{A}}^2(z^+)\; ,
\end{equation}
by the same symbol, but without an argument.

\subsection{Non-tadpole corrections}
We can now write
\begin{equation}
{\cal M}={\cal T}
{\cal N}\; ,
\label{eq:TN}
\end{equation}
where ${\cal N}$ is the same as ${\cal M}$, but containing only links
that connect different Wilson lines. A typical term in ${\cal N}$ with
two links could be the following:
\setbox1=%
\hbox to 2.4cm{\hfil\resizebox*{2.4cm}{!}{\includegraphics{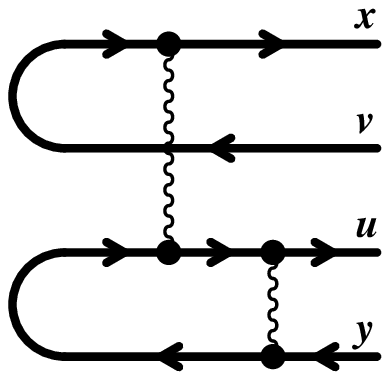}}}
\begin{equation}
\raise -10mm\box1\;\;\; .
\label{eq:example-N}
\end{equation}
Note that since $\left<\rho_{_{A}} \rho_{_{A}}\right>$ is local in $x^+$, links
must connect points at the same $x^+$. Therefore, diagrams with crossed
links are not allowed.  By a systematic use of the Fierz identity of
eq.~(\ref{eq:fierz}), we can completely remove the links. On the
previous example, this would lead to the following four
terms\footnote{The diagrams have been rearranged a bit in order to make
  them look nicer, without changing their topology.}:
\setbox1=%
\hbox to 1.6cm{\hfil\resizebox*{1.6cm}{!}{\includegraphics{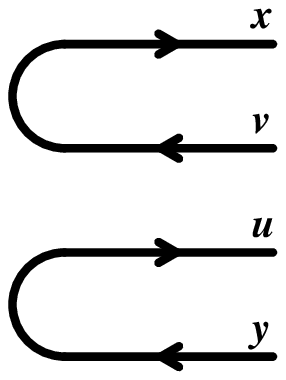}}}
\setbox2=%
\hbox to 1.6cm{\hfil\resizebox*{1.6cm}{!}{\includegraphics{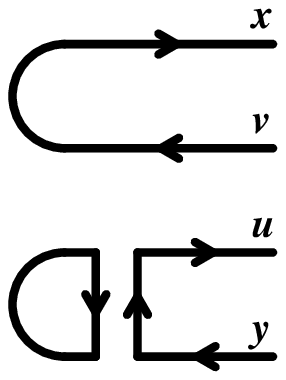}}}
\setbox3=%
\hbox to 1.6cm{\hfil\resizebox*{1.6cm}{!}{\includegraphics{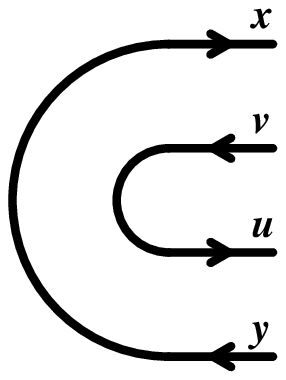}}}
\setbox4=%
\hbox to 1.6cm{\hfil\resizebox*{1.6cm}{!}{\includegraphics{fig_in_text8a.ps}}}
\begin{equation}
\frac{1}{4}\;\raise -10mm\box1
\;-\frac{1}{4N}\;\raise -10mm\box2
\;-\frac{1}{4N}\;\raise -10mm\box3
\;+\frac{1}{4N^2}\;\raise -10mm\box4\;\;\; .
\end{equation}
Note that the closed loop that appears in the second term can be
evaluated directly and replaced by the trace of the unit matrix in the
fundamental representation ${\rm tr}_{\rm c}(1_{_{F}})=N$. Therefore,
we see that all the terms that are generated in the process of
eliminating the links with the Fierz identity are one of the following
two terms:
\setbox1=%
\hbox to 1.6cm{\hfil\resizebox*{1.6cm}{!}{\includegraphics{fig_in_text8a.ps}}}
\setbox2=%
\hbox to 1.6cm{\hfil\resizebox*{1.6cm}{!}{\includegraphics{fig_in_text8c.ps}}}
\begin{equation}
{\cal N}^{(a)}\equiv\; \raise -10mm\box1\quad\; \qquad
{\cal N}^{(b)}\equiv\; \raise -10mm\box2\;\;\; .
\end{equation}
This property, observed for the example of eq.~(\ref{eq:example-N}), is
in fact completely general.

Let us now denote ${\cal N}_n(z_1^+,\cdots,z_n^+)$ the piece of ${\cal
  N}$ that contains $n$ links, at the ``times'' $z_1^+<\cdots<z_n^+$.
We therefore have, 
\begin{equation}
{\cal N}=\sum_{n=0}^{+\infty}\;
\int\limits_{z_1^+<\cdots<z_n^+}\!\!\!\!\!\!
{\cal N}_n(z_1^+,\cdots,z_n^+)
\; .
\label{eq:N-def}
\end{equation}
Using the above remarks, we can write each ${\cal N}_n$ as a linear combination 
of ${\cal N}^{(a)}$ and ${\cal N}^{(b)}$:
\begin{equation}
{\cal N}_n\equiv a_n {\cal N}^{(a)} + b_n {\cal N}^{(b)}\; .
\label{eq:Nn-decomp}
\end{equation}
Note that the term with zero links is exactly ${\cal N}^{(a)}$, so
that we have $a_0=1$ and $b_0=0$. Once we know ${\cal N}_{n-1}$, the
next term in the expansion, ${\cal N}_{n}$, can be obtained by adding
one link to ${\cal N}^{(a)}$ and ${\cal N}^{(b)}$ in all the possible
ways:
\setbox1=%
\hbox to 1.2cm{\hfil\resizebox*{1.2cm}{!}{\includegraphics{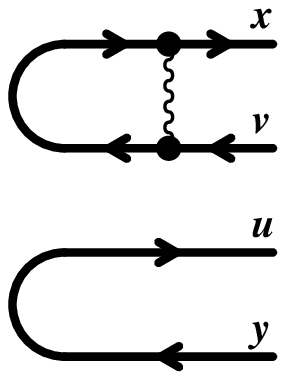}}}
\setbox2=%
\hbox to 1.2cm{\hfil\resizebox*{1.2cm}{!}{\includegraphics{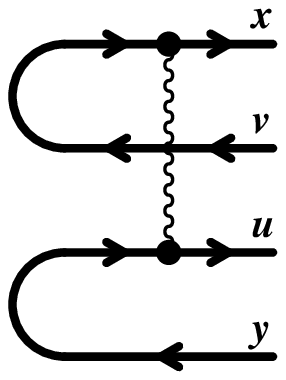}}}
\setbox3=%
\hbox to 1.2cm{\hfil\resizebox*{1.2cm}{!}{\includegraphics{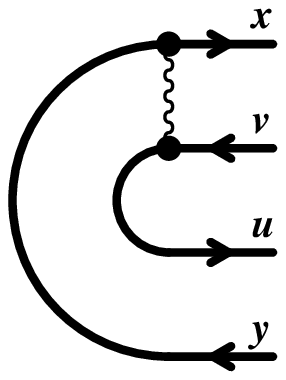}}}
\setbox4=%
\hbox to 1.2cm{\hfil\resizebox*{1.2cm}{!}{\includegraphics{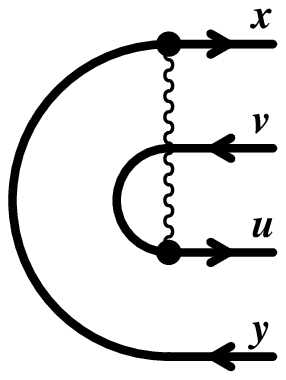}}}
\begin{eqnarray}
{\cal N}_{n}&=&
\mu_{_{A}}^2(z_n^+)\,a_{n-1}
\Bigg[
L(x,v)\;\raise -7mm\box1
\;-\;
L(x,u)\;\raise -7mm\box2
\;+\cdots
\Bigg]\nonumber\\
&+&
\mu_{_{A}}^2(z_n^+)\,b_{n-1}
\Bigg[
L(x,v)\;\raise -7mm\box3
\;-\;
L(x,u)\;\raise -7mm\box4
\;+\cdots
\Bigg]\; ,
\end{eqnarray}
where, generalizing eq.~(\ref{eq:L-1}), we denote
\begin{equation}
L(x,v)\equiv g^4
\int_{\z_\perp}G_0(\x_\perp-\z_\perp)G_0(\v_\perp-\z_\perp)\; .
\label{eq:L-2}
\end{equation}
There are six possible ways to link four lines pairwise, and we have
only represented two of them. Note that there is a $-$ sign for links
connecting two Wilson lines going in the same direction (i.e. two
${\wt U}$'s or two ${\wt U}^\dagger$'s). The next step is to use
eq.~(\ref{eq:fierz}) in order to remove all the links and to replace
by $N$ the closed loops that may appear in the process. Finally,
identifying with ${\cal N}_{n}=a_{n}{\cal N}^{(a)}+b_{n} {\cal
  N}^{(b)}$, we obtain the following recursion relation:
\begin{equation}
\genfrac{(}{)}{0pt}{}{a_{n}}{b_{n}}
=
\mu_{_{A}}^2(z_{n}^+) {\bs M}
\genfrac{(}{)}{0pt}{}{a_{n-1}}{b_{n-1}}\; ,
\label{eq:recursion-1}
\end{equation}
where ${\bs M}$ is a $2\times 2$ matrix given by
\begin{equation}
{\bs M}\equiv
\begin{pmatrix}
\alpha C_{_{F}}+\frac{1}{2N}(\beta-\gamma)
&
\frac{1}{2} (\alpha-\beta)
\\
\frac{1}{2}(\gamma-\beta)
&
\gamma C_{_{F}}+\frac{1}{2N}(\beta-\alpha)
\\
\end{pmatrix}
\; ,
\end{equation}
with
\begin{eqnarray}
&&\alpha\equiv L(x,v)+L(u,y)\; ,\nonumber\\
&&\beta\equiv  L(x,u)+L(v,y)\; ,\nonumber\\
&&\gamma\equiv L(x,y)+L(v,u)\; .
\label{eq:recursion-2}
\end{eqnarray}
Note that ${\bs M}$ does not depend on $z_{n}^+$. Given the initial
condition $a_0=1, b_0=0$, it is now very easy to formally solve the
recursion relation to obtain
\begin{equation}
\genfrac{(}{)}{0pt}{}{a_{n}}{b_{n}}
=
\left[\prod_{i=1}^n \mu_{_{A}}^2(z_i^+)\right]
{\bs M}^n 
\genfrac{(}{)}{0pt}{}{1}{0}\; .
\label{eq:recursion-3}
\end{equation}

From the coefficients $a_n, b_n$, the correlator
$C(\x_\perp,\y_\perp;\u_\perp,\v_\perp)$ is then obtained by using
eqs.~(\ref{eq:C-decomp}), (\ref{eq:TN}), (\ref{eq:N-def}) and
(\ref{eq:Nn-decomp}). Using eq.~(\ref{eq:C-decomp}) amounts to ``closing
the ends'' of ${\cal N}$. Multiplying also by the contribution of the
tadpole insertions, we get:
\begin{equation}
C(\x_\perp,\y_\perp;\u_\perp,\v_\perp)=\frac{1}{2}
\left[
\sum_{n=0}^{+\infty}\;
\int\limits_{z_1^+<\cdots<z_n^+}
\!\!\!\!\!\!a_n
\right]
\Big(N^2-1\Big){\cal T}
\; .
\label{eq:C-tmp}
\end{equation}
In order to make this formula more explicit, we must integrate
eq.~(\ref{eq:recursion-3}) over the $z_i^+$'s, which
gives\footnote{The ordering of the $z_i^+$'s in eq.~(\ref{eq:N-def})
  can be trivially removed because the recursion matrix commutes with
  itself at different times. This gives a factor $1/n!$.}:
\begin{equation}
\int\limits_{z_1^+<\cdots<z_n^+}
\genfrac{(}{)}{0pt}{}{a_{n}}{b_{n}}
=\frac{1}{n!}
\mu_{_{A}}^{2n}
{\bs M}^n 
\genfrac{(}{)}{0pt}{}{1}{0}\; .
\end{equation}
The only remaining task is to calculate the $n$-th power of the
constant matrix ${\bs M}$. This  can be done by finding its
eigenvalues, which are easily shown to be 
\begin{equation}
\lambda_\pm=\frac{1}{2}\left[
\frac{N}{2}(\alpha+\gamma)
+
\frac{1}{N}(\beta-\alpha-\gamma)
\pm\sqrt{
\frac{N^2}{4}(\alpha-\gamma)^2+(\alpha-\beta)(\gamma-\beta)
}
\right]\; .
\end{equation}
It is then clear that we can write
\begin{equation}
\sum_{n=0}^{+\infty}\;
\int\limits_{z_1^+<\cdots<z_n^+}\!\!\!\!\!\!a_n
= a_+ e^{\mu_{_{A}}^2 \lambda_+} + a_- e^{\mu_{_{A}}^2 \lambda_-} \; ,
\end{equation}
with $a_+$ and $a_-$ two numbers determined from the initial
condition. At this point, it is a simple matter of algebra to derive
our final expression for the 4-point function $C$:
\begin{eqnarray}
C(\x_\perp,\y_\perp;\u_\perp,\v_\perp)
\!\!\!&=&\!\!\!NC_{_{F}}
e^{
-\frac{1}{4N}\mu_{_{A}}^2\big[
\Gamma(x-u)+\Gamma(y-v)
\big]
}
\nonumber\\
&&\!\!\!\!\!\times
e^{-\big(\frac{N}{8}-\frac{1}{4N}\big)\mu_{_{A}}^2
\big[
\Gamma(x-v)+\Gamma(x-y)+\Gamma(u-v)+\Gamma(u-y)
\big]}\nonumber\\
&&\!\!\!\!\!\times\Bigg\{
\frac{1+\sqrt{\Delta}}{2\sqrt{\Delta}}
e^{+\frac{N}{4}\mu_{_{A}}^2(\alpha-\gamma)\sqrt{\Delta}}
-\frac{1-\sqrt{\Delta}}{2\sqrt{\Delta}}
e^{-\frac{N}{4}\mu_{_{A}}^2(\alpha-\gamma)\sqrt{\Delta}}
\Bigg\}\; ,\nonumber\\
&&
\label{eq:C-final}
\end{eqnarray}
where we denote
\begin{equation}
\Delta\equiv 1+\frac{16}{N^2}
\frac{(\alpha-\beta)(\gamma-\beta)}{(\alpha-\gamma)^2}
\; ,
\end{equation}
and
\begin{equation}
\Gamma(y-u)\equiv L(y,y)+L(u,u)-2L(y,u)\; .
\label{eq:Gamma}
\end{equation}
A comment is in order regarding this result. The quantity $L(x,y)$
defined in eq.~(\ref{eq:L-2}) is strongly infrared divergent. Indeed,
if one introduces by hand an upper cutoff in the integration over
$\z_\perp$ at $z_\perp \sim \Lambda_{_{QCD}}^{-1}$, then $L(x,y)\sim
\Lambda_{_{QCD}}^{-2}$. However, the combination $\Gamma$ of these
objects, defined in eq.~(\ref{eq:Gamma}), has a much milder infrared
singularity which is logarithmic in the cutoff scale. This is also the
case of the differences $\alpha-\gamma$, $\alpha-\beta$ and
$\gamma-\beta$ which appear in eq.~(\ref{eq:C-final}). Therefore, we
conclude that the 4-point function studied here is not more infrared
singular than the 2-point functions already discussed extensively in
the literature. Physically, the reason why the quadratic divergence is
softened to a logarithmic divergence is that we consider a correlator
made of a singlet combination of Wilson lines.  The reason why we have
nevertheless a residual logarithmic divergence is that the color
neutrality of the nucleus is not enforced in the functional of
eq.~(\ref{eq:W}). We have introduced by hand a cutoff at the scale
$\Lambda_{_{QCD}}$ which is equivalent to imposing color
neutralization at the scale of the size of the nucleon.

\subsection{Large $N$ limit}
Eq.~(\ref{eq:C-final}) becomes substantially simpler in the large $N$
limit. Indeed, in this limit, we have $\Delta=1$. We have therefore, 
\begin{equation}
C(\x_\perp,\y_\perp;\u_\perp,\v_\perp)\empile{=}\over{N\to \infty}
\frac{N^2}{2}
e^{-\frac{N}{8} \mu_{_{A}}^2
\big[
\Gamma(x-v)+\Gamma(x-y)+\Gamma(u-v)+\Gamma(u-y)
\big]
}
e^{\frac{N}{4}\mu_{_{A}}^2(\alpha-\gamma)}\, .
\end{equation}
Using the explicit expressions of $\alpha,\gamma$
(eq.~(\ref{eq:recursion-2})), this simplifies greatly to read, 
\begin{equation}
C(\x_\perp,\y_\perp;\u_\perp,\v_\perp)\empile{=}\over{N\to \infty}
\frac{N^2}{2}
e^{-\frac{N}{4}\mu_{_{A}}^2\left[\Gamma(x-v)
+\Gamma(y-u)\right]}\; .
\end{equation}
This result could also have been obtained very simply from the
recursion relation eq.~(\ref{eq:recursion-1}) because the recursion
matrix ${\bs M}$ is diagonal in the large $N$ limit.

\subsection{Special cases}
There are a few degenerate cases in which the 4-point function we have
calculated in this appendix simplifies into a simpler function. These
special cases are useful as a check, because for these the result is
already known in the literature.

\subsubsection{$\x_\perp=\v_\perp$ or $\y_\perp=\u_\perp$}
There is an obvious simplification in
$C(\x_\perp,\y_\perp;\u_\perp,\v_\perp)$ if $\x_\perp=\v_\perp$ or if
$\y_\perp=\u_\perp$.  Indeed, any of these equalities makes
$\beta=\gamma$, which implies $\Delta=1$. For instance, in the case of
$\x_\perp=\v_\perp$, we obtain from eq.~(\ref{eq:C-final}) the
following formula:
\begin{eqnarray}
C(\x_\perp,\y_\perp;\u_\perp,\x_\perp)
\!\!\!&=&\!\!\!NC_{_{F}}
e^{
-\frac{1}{4N}\mu_{_{A}}^2\big[
\Gamma(x-u)+\Gamma(x-y)
\big]
}
\nonumber\\
&&\!\!\!\times
e^{-\big(\frac{N}{8}-\frac{1}{4N}\big)\mu_{_{A}}^2
\big[
\Gamma(x-y)+\Gamma(x-u)+\Gamma(y-u)
\big]}
e^{\frac{N}{4}\mu_{_{A}}^2(\alpha-\gamma)}\nonumber\\
&=&NC_{_{F}} e^{
-\frac{C_{_{F}}}{2}\mu_{_{A}}^2\Gamma(y-u)
}\; .
\end{eqnarray}
This simple formula is indeed what we already know (see \cite{GelisP1}
for instance) for this degenerate case, since we expect:
\begin{equation}
C(\x_\perp,\y_\perp;\u_\perp,\x_\perp)\!=\!
C_{_{F}}{\rm tr}_{\rm c}\left<
{\wt U}^\dagger(\y_\perp) {\wt U}(\u_\perp)
\right>
\, .
\end{equation}

\subsubsection{$\x_\perp=\y_\perp$ or $\u_\perp=\v_\perp$}
A slightly less trivial simplification is obtained when
$\x_\perp=\y_\perp$ or $\u_\perp=\v_\perp$. For instance if
$\u_\perp=\v_\perp$, the correlator $C$ should become the 3-point
function ${\rm tr}_{\rm c}\big<{\wt U}(\x_\perp)t^a {\wt
U}^\dagger(\y_\perp)t^b U^{ba}(\u_\perp)\big>$, whose expression can
be found in the appendix of a paper by Kovner and Wiedemann
\cite{KovneW1}. In the case where one of these two equalities is
satisfied, we have $\alpha=\beta$, which also implies $\Delta=1$. When
$\u_\perp=\v_\perp$, eq.~(\ref{eq:C-final}) simplifies into:
\begin{eqnarray}
C(\x_\perp,\y_\perp;\u_\perp,\u_\perp)
\!\!\!&=&\!\!\!NC_{_{F}}
e^{
-\frac{1}{4N}\mu_{_{A}}^2\big[
\Gamma(x-u)+\Gamma(y-u)
\big]
}
\nonumber\\
&&\!\!\!\times
e^{-\big(\frac{N}{8}-\frac{1}{4N}\big)\mu_{_{A}}^2
\big[
\Gamma(x-u)+\Gamma(x-y)+\Gamma(u-y)
\big]}
e^{\frac{N}{4}\mu_{_{A}}^2(\alpha-\gamma)}
\nonumber\\
&=&NC_{_{F}}\,e^{
-\frac{N}{4}\mu_{_{A}}^2\big[\Gamma(x-u)+\Gamma(y-u)\big]
-\left(\frac{C_{_{F}}}{2}-\frac{N}{4}\right)\mu_{_{A}}^2\Gamma(x-y)
}\, .
\end{eqnarray}
It is trivial to check that this formula is equivalent to eq.~(C7) of
\cite{KovneW1} (the correspondence between our $\Gamma(x-y)$ and the
$v(x-y)$ defined in \cite{KovneW1} is $2v(x-y)=\mu_{_{A}}^2\Gamma(x-y)$).

\section{Retarded pair production amplitude}
\label{sec:ret-amp}
\subsection{Introduction}
In section \ref{sec:F-amplitude}, we have calculated the time-ordered
pair production amplitude, which is directly related to the
probability $P_1$ of producing a single quark-antiquark pair in a
collision. We also conjectured that, since we are treating the problem
at leading order in the source $\rho_p$ that describes the proton, at
most one pair can be produced in a collision. This implies that the
average number of pairs $\overline{N}_{q\bar{q}}$ produced in a
collision must be equal to the probability $P_1$, simply because the
probabilities $P_2$, $P_3, \cdots$ to produce 2 or more pairs in a
collision are all zero at this order.

On the other hand, we can calculate directly the average pair
multiplicity $\overline{N}_{q\bar{q}}$ from the retarded pair
production amplitude, as explained in \cite{BaltzGMP1}, from the formula,
\begin{equation}
\overline{N}_{q\bar{q}}=
\int\frac{d^3\q}{(2\pi)^3 2E_\q}
\int\frac{d^3\p}{(2\pi)^3 2E_\p}
\left|{\cal M}_{_{R}}(\q,\p)\right|^2\; ,
\end{equation}
where ${\cal M}_{_{R}}(\q,\p)$ is the pair production amplitude
calculated with retarded quark propagators (as opposed to Feynman
quark propagators in the case of ${\cal M}_{_{F}}(\q,\p)$).  Since the
calculation of the retarded amplitude presents a number of significant
differences with respect to the time-ordered amplitude, its
calculation provides a non-trivial cross-check of the results we have
derived in section \ref{sec:F-amplitude}. Because of the expected
equality $P_1=\overline{N}_{q\bar{q}}$, we expect the retarded and
time-ordered amplitude to be identical ``up to a phase''.

\begin{figure}[htbp]
\begin{center}
\resizebox*{!}{4cm}{\includegraphics{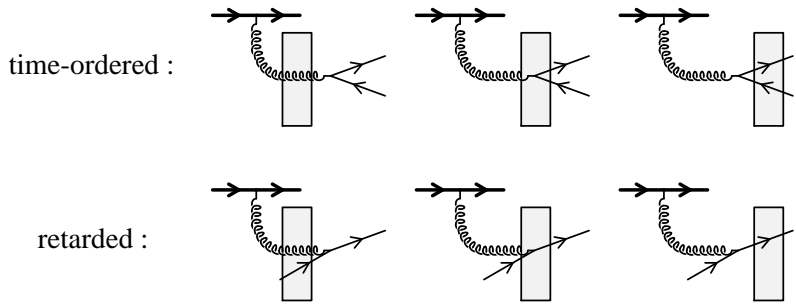}}
\end{center}
\caption{\label{fig:retarded} Top: the contributions to the Feynman
amplitude.  The shaded area represents the nucleus. A gluon is emitted
by the proton, and splits into a pair: after the collisions with the
nucleus (left), inside the nucleus (middle) or before the collision
(right). Bottom: the same contributions for the retarded amplitude.}
\end{figure}
In fact, without doing any calculation, we can guess what this phase
is. The contributions to the two amplitudes are sketched in figure
\ref{fig:retarded}, and one can see that the only difference between
the two is that the antiquark line comes from the region of negative
$x^+$ in the case of the retarded amplitude. This is so because the
retarded quark propagator can only move forward in time.  From this
figure, we see that moving the antiquark line from the left of the
diagram to the right of the diagram changes the amplitude by a factor
which is precisely the scattering matrix of the antiquark on the
nucleus. In a sense to be made more precise later, we expect the
following relation:
\begin{equation}
{\cal M}_{_{R}}={\cal M}_{_{F}}\; {\wt U}(\y_\perp)\; ,
\end{equation}
where $\y_\perp$ is the transverse coordinate of the antiquark.

In order to calculate explicitly the retarded amplitude, we must use
the free retarded quark propagator:
\begin{equation}
S_{_{R}}^0(p)\equiv i\frac{\slp+m}{p^2-m^2+ip^+\epsilon}\; ,
\label{eq:Sr}
\end{equation}
which differs from eq.~(\ref{eq:Sf}) by the location of its poles in
the complex energy plane. Note that the $p^+\epsilon$ in the
denominator could as well have been $p^-\epsilon$ or $p^0\epsilon$
because $p^+, p^-$ and $p^0$ have the same sign at the pole. The other
technical difference is the form of the scattering matrix of a
quark/antiquark on the nucleus. In the retarded case, it is given by
\cite{BaltzGMP1}: 
\setbox1=%
\hbox to 1.3cm{\resizebox*{1.3cm}{!}{\includegraphics{fig_in_text1.ps}}}
\begin{equation}
\raise -4mm\box1=2\pi\delta(k^+)\gamma^+\,
\int d^2\x_\perp e^{i\k_\perp\cdot\x_\perp}
\left[
{\wt U}(\x_\perp)-1
\right]\; .
\label{eq:R-scatt}
\end{equation}
We see that it coincides with the time-ordered scattering operator
given in eq.~(\ref{eq:F-scatt}) when $p^+>0$ (as expected since the
poles of the retarded and time-ordered free propagator are the same if
$p^+>0$), but differs from it when $p^+<0$.

\subsection{Singular terms}
Let us start with the calculation of the term involving the field
$A_{\rm sing}^\mu$, where the pair is produced inside the
nucleus. It is trivial to reproduce the calculation of section
\ref{sec:F-singular-term} with the Feynman rules appropriate for the
retarded propagator. We obtain:
\begin{eqnarray}
&&{\cal M}_{_{R}}^{\rm sing}(\q,\p)=g^2\int\frac{d^2\k_{1\perp}}{(2\pi)^2}
\frac{\rho_{p,a}(\k_{1\perp})}{k_{1\perp}^2}
\int d^2\x_\perp
e^{i(\p_\perp+\q_\perp-\k_{1\perp})\cdot\x_\perp}
\nonumber\\
&&\qquad\qquad\qquad
\times
\frac{\overline{u}(\q)\gamma^+ t^b {\wt U}(\x_\perp)v(\p)}{p^++q^+}
\big[
V(\x_\perp)-U(\x_\perp)
\big]_{ba}\; .
\label{eq:Mr-sing}
\end{eqnarray}
This is indeed the same as eq.~(\ref{eq:Mf-sing}), up to
the factor ${\wt U}(\x_\perp)$ (the transverse coordinate of the
antiquark is $\x_\perp$ in this expression).

\subsection{Regular terms}
We must now calculate the four ``regular'' (in the sense that the pair
is produced outside of the nucleus) diagrams of figure
\ref{fig:reg-diagrams} in the case of the retarded amplitude.  The
diagram (a) does not contain any internal quark propagator and is
therefore identical in the time-ordered and retarded cases:
\begin{equation}
{\cal M}_{_{R}}^{(a)}(\q,\p)={\cal M}_{_{F}}^{(a)}(\q,\p)\; .
\end{equation}
Contrary to the time-ordered case, the pole structure of the diagram
(d) is such that it does not contribute in the retarded
case\footnote{This property is a consequence of
causality. Indeed, in the retarded case, the diagram (d) corresponds
to the following process: a quark goes through the nucleus, absorbs
the gluon coming from the proton, and goes again through the
nucleus. At high collision energy, this is suppressed by one power of
the collision energy.}:
\begin{equation}
{\cal M}_{_{R}}^{(d)}(\q,\p)=0\; .
\end{equation}
The diagram (c) is also very simple: indeed, the only fermion
propagator that interacts with the nucleus is the quark line which
carries the momentum $q^+>0$. Therefore, this propagator is the same
in the retarded and in the time-ordered case:
\begin{equation}
{\cal M}_{_{R}}^{(c)}(\q,\p)={\cal M}_{_{F}}^{(c)}(\q,\p)\; .
\end{equation}
All the difficulties are concentrated in the diagram (b). Indeed,
since this diagram has a fermion propagator carrying the momentum
$-p^+<0$, its pole structure is not the same in the retarded and in
the time-ordered case. In addition, the interaction operator of the
antiquark on the nucleus is different. The diagram (b) can be divided
in three terms:
\begin{equation}
{\cal M}_{_{R}}^{(b)}=
{\cal M}_{_{R},1}^{(b)}
+
{\cal M}_{_{R},2}^{(b)}
+
{\cal M}_{_{R},3}^{(b)}\; .
\end{equation}
${\cal M}_{_{R},1}^{(b)}$ is obtained by keeping only the first line
of the field in eq.~(\ref{eq:field-1}) (namely replacing the
solution of Yang-Mills equations by the field of the proton
alone). This term can be written as follows:
\begin{eqnarray}
&&\!\!\!\!\!\!\!\!\!\!\!\!
{\cal M}_{_{R},1}^{(b)}(\q,\p)\!=\!g^2\!\!
\int\!\frac{d^2\k_{1\perp}}{(2\pi)^2}
\frac{d^2\k_\perp}{(2\pi)^2}
\frac{\rho_{p,a}(\k_{1\perp})}{k_{1\perp}^2}
\!\!\int\!\! d^2\x_\perp d^2\y_\perp
e^{i\k_\perp\cdot\x_\perp}
e^{i(\p_\perp\!+\!\q_\perp\!-\!\k_\perp\!-\!\k_{1\perp})\cdot\y_\perp}
\nonumber\\
&&\times
\frac{\overline{u}(\q)\gamma^+(\slq\!-\!\slk+m)\gamma^- (\slq\!-\!\slk\!-\!\slk_1+m)
\gamma^+ 
[t^a\!-\!t^a {\wt U}(\y_\perp)]
\!v(\p)}
{2p^+((\q_\perp-\k_\perp)^2+m^2)
+2q^+((\q_\perp-\k_\perp-\k_{1\perp})^2+m^2)}
\; .
\label{eq:Mr-b-1}
\end{eqnarray}
The term we denote ${\cal M}_{_{R},2}^{(b)}$ is obtained from the
diagram (b) by keeping only the piece proportional to $C_{_{V},{\rm
reg}}^\mu$ in eq.~(\ref{eq:field-1}). Using the fact that
\begin{equation}
C_{_{V},{\rm reg}}^\mu(p+q-k)=2(p+q-k)^\mu 
- \delta^{\mu -}\frac{(p+q-k)^2}{p^++q^+-k^+}\; ,
\end{equation}
we obtain:
\begin{eqnarray}
&&\!\!\!\!\!\!
\int\frac{dk^-}{2\pi}
\frac{\overline{u}(\q)t^b \slC_{_{V},{\rm reg}}(p+q-k)(-\slp+\slk+m)
\gamma^+ [{\wt U}(\x_\perp)-1]v(\p)}
{[(p+q-k)^2+i(p^++q^+-k^+)\epsilon][(-p+k)^2-m^2+i(-p^++k^+)\epsilon]}
\nonumber\\
&&\qquad\qquad
=i\frac{\overline{u}(\q)\gamma^+ t^b [{\wt U}(\x_\perp)-1]v(\p)}
{p^++q^+}\; .
\end{eqnarray}
With the help of this relation, we obtain easily:
\begin{eqnarray}
&&\!\!\!\!{\cal M}_{_{R},2}^{(b)}=-g^2
\int \frac{d^2\k_{1\perp}}{(2\pi)^2}
\frac{\rho_{p,a}(\k_{1\perp})}{k_{1\perp}^2}
\int d^2\x_\perp 
e^{i(\p_\perp+\q_\perp-\k_{1\perp})\cdot\x_\perp}\nonumber\\
&&\times
\frac{\overline{u}(\q)\gamma^+ t^b [{\wt U}(\x_\perp)-1]v(\p)}
{p^++q^+}
[V(\x_\perp)-1]_{ba}\; .
\label{eq:Mr-b-2}
\end{eqnarray}
Finally, ${\cal M}_{_{R},3}^{(b)}$ is obtained from the diagram (b) by
keeping only the piece proportional to $C_{_{U}}^\mu$ in
eq.~(\ref{eq:field-1}). It reads:
\begin{eqnarray}
&&\!\!\!\!\!{\cal M}_{_{R},3}^{(b)}=ig^2\!\!
\int \!\!\frac{d^2\k_{1\perp}}{(2\pi)^2}
\frac{d^2\k_\perp}{(2\pi)^2}
\frac{\rho_{p,a}(\k_{1\perp})}{k_{1\perp}^2}
\!\!\int\!\! d^2\x_\perp d^2\y_\perp
e^{i\k_\perp\cdot\y_\perp}
e^{i(\p_\perp+\q_\perp-\k_\perp-\k_{1\perp})\cdot\x_\perp}\nonumber\\
&&\!\!\!\!\!\times\!\!
\int\!\!\frac{dk^-}{2\pi}
\frac{\overline{u}(\q)\slC_{_{U}}(p\!+\!q\!-\!k,\k_{1\perp}\!)
(\slk\!-\!\slp\!+\!m)
\gamma^+ t^b [{\wt U}(\y_\perp)\!-\!1]v(\p)}
{
[(p+q-k)^2+i\epsilon][(-p+k)^2-m^2-i\epsilon]
}[U(\x_\perp)\!-\!1]_{ba}\, .
\nonumber\\
&&
\end{eqnarray}

\subsection{Total retarded amplitude}
\label{sec:MR-regular}
The main difficulty at this point is that we must combine this term in
$C_{_{U}}^\mu(p+q-k,\k_{1\perp})$ with a term in
$C_{_{U}}^\mu(p+q,\k_{1\perp})$ contained in ${\cal M}_{_{R}}^{(a)}$. In
order to do this, one must notice the following relation:
\begin{eqnarray}
&&\frac{\overline{u}(\q)t^b \slC_{_{U}}(p+q,\k_{1\perp})v(\p)}{(p+q)^2}=
\nonumber\\
&&\quad=\left.
i\!\int\frac{dk^-}{2\pi}
\frac{\overline{u}(\q)t^b \slC_{_{U}}(p+q-k,\k_{1\perp})(-\slp+\slk+m)
\gamma^+v(\p)}{
[(p+q-k)^2+i\epsilon][(-p+k)^2-m^2-i\epsilon]
}
\right|_{\k_\perp=0}
\; ,
\label{eq:Cu-trick}
\end{eqnarray}
which is valid if $\k_\perp=0$ in the r.h.s. and introduce in ${\cal
M}_{_{R}}^{(a)}$ dummy variables $\y_\perp$ and $\k_\perp$ via a
trivial factor (which ensures that $\k_\perp=0$):
\begin{equation}
1=\int \frac{d^2\k_\perp}{(2\pi)^2} d^2\y_\perp e^{i\k_\perp\cdot
\y_\perp}\; .
\end{equation}
In order to simplify the notations, let us denote:
\begin{equation}
{\cal F}(\q_\perp,\p_\perp-\k_\perp,\k_{1\perp}) \equiv
i\int\frac{dk^-}{2\pi}
\frac{\slC_{_{U}}(p+q-k,\k_{1\perp})(-\slp+\slk+m)}{[(p+q-k)^2+i\epsilon][(-p+k)^2-m^2-i\epsilon]}\;
,
\label{eq:F-def}
\end{equation}
where we have only indicated the transverse momenta in the list of
arguments. Note that this object depends only on the difference
$\p_\perp-\k_\perp$ rather than $\p_\perp$ and $\k_\perp$ separately.
Finally, we need to perform the transformation $\k_\perp \to
\p_\perp+\q_\perp-\k_{1\perp}-\k_\perp$ for the terms containing
$C_{_{U}}$ (or equivalently the ${\cal F}$ defined above). Combining
all the contributions to the retarded pair production amplitude, we
can write it as follows:
\begin{eqnarray}
{\cal M}_{_{R}}(\q,\p)=\int d^2\y_\perp e^{i\p_\perp\cdot\y_\perp}
\overline{u}(\q) {\cal K}(\q_\perp;\y_\perp) {\wt U}(\y_\perp)\gamma^+ v(\p)\; ,
\label{eq:M-R}
\end{eqnarray}
with
\begin{eqnarray}
&&{\cal K}(\q_\perp;\y_\perp)\equiv g^2
\int \frac{d^2\k_{1\perp}}{(2\pi)^2}
\frac{d^2\k_\perp}{(2\pi)^2}
\frac{\rho_{p,a}(\k_{1\perp})}{k_{1\perp}^2}
\int d^2\x_\perp
e^{i\k_\perp\cdot\x_\perp}
e^{i(\q_\perp-\k_\perp-\k_{1\perp})\cdot\y_\perp}\nonumber\\
&&\quad
\times\Bigg\{
\frac{\gamma^+(\slq\!-\!\slk+m)\gamma^- (\slq\!-\!\slk\!-\!\slk_1+m)
\left[{\wt U}(\x_\perp)t^a
{\wt U}^\dagger(\y_\perp)\!-\!t^a\right]}
{2p^+((\q_\perp-\k_\perp)^2+m^2)
+2q^+((\q_\perp-\k_\perp-\k_{1\perp})^2+m^2)}\nonumber\\
&&\qquad+\Bigg[t^b 
{\cal F}(\q_\perp,\k_\perp+\k_{1\perp}-\q_\perp,\k_{1\perp})-
\frac{t^b}{p^++q^+}
\Bigg]
[U(\x_\perp)-1]_{ba}
\Bigg\}\; .
\label{eq:K-def}
\end{eqnarray}
Note that the quantity ${\cal K}(\q_\perp;\y_\perp)$ does not depend
on $\p_\perp$.

\subsection{Comparison with the time-ordered amplitude}
Let us now go back to our original problem, namely to verify that we
indeed have $P_1=\overline{N}_{q\bar{q}}$. In order to make this
verification easy, the time-ordered amplitude given in
eq.~(\ref{eq:Mf-final}) needs some rewriting. By performing on the
term in $C_{_{U}}$ the transformations described in the section
\ref{sec:MR-regular}, we can write:
\begin{eqnarray}
{\cal M}_{_{F}}(\q,\p)=\int d^2\y_\perp e^{i\p_\perp\cdot\y_\perp}
\overline{u}(\q) {\cal K}(\q_\perp;\y_\perp) \gamma^+ v(\p)\; ,
\label{eq:M-F}
\end{eqnarray}
with the same ${\cal K}(\q_\perp;\y_\perp)$ as in
eq.~(\ref{eq:K-def}). We see that the time-ordered and the retarded
amplitudes differ only by the unitary matrix ${\wt U}(\y_\perp)$ in
the integrand: this is the precise meaning of ``identical up to a
phase'' in this context.

From eq.~(\ref{eq:M-F}), we can write the single quark spectrum as:
\begin{eqnarray}
&&\frac{d P_1}{d^2\q_\perp dy_q}=\frac{1}{64\pi^4}
\int\frac{d^2\p_\perp}{(2\pi)^2}\frac{dp^+}{p^+}
\int d^2\y_\perp d^2\y_\perp^\prime
e^{i\p_\perp\cdot(\y_\perp-\y_\perp^\prime)}
\nonumber\\
&&\qquad\times
{\rm tr}_{\rm d}\big[
(\slq+m){\cal K}(\q_\perp;\y_\perp)
\gamma^+(\slp-m)\gamma^+
\gamma^0 {\cal K}^\dagger(\q_\perp,\y_\perp^\prime)\gamma^0
\big]\; .
\end{eqnarray}
Since $\gamma^+(\slp-m)\gamma^+=2 p^+ \gamma^+$ does not depend on
$\p_\perp$, nor does ${\cal K}$, it is trivial to perform the
integration over $\p_\perp$, which gives a
$\delta(\y_\perp-\y_\perp^\prime)$. Therefore, we can rewrite the
above equation as follows:
\begin{equation}
\frac{d P_1}{d^2\q_\perp dy_q}=\frac{1}{32\pi^4}
\int{dp^+} d^2\y_\perp 
{\rm tr}_{\rm d}\big[
(\slq+m){\cal K}(\q_\perp;\y_\perp)
\gamma^+
\gamma^0 {\cal K}^\dagger(\q_\perp,\y_\perp)\gamma^0
\big]\; .
\end{equation}
Similarly, one can write the average number of produced quarks as:
\begin{eqnarray}
&&\frac{d \overline{N}_{q\bar{q}}}{d^2\q_\perp dy_q}=\frac{1}{64\pi^4}
\int\frac{d^2\p_\perp}{(2\pi)^2}\frac{dp^+}{p^+}
\int d^2\y_\perp d^2\y_\perp^\prime
e^{i\p_\perp\cdot(\y_\perp-\y_\perp^\prime)}
\nonumber\\
&&\quad\times
{\rm tr}_{\rm d}\big[
(\slq+m){\cal K}(\q_\perp;\y_\perp)
{\wt U}(\y_\perp)\gamma^+(\slp-m)\gamma^+{\wt U}^\dagger(\y_\perp^\prime)
\gamma^0 {\cal K}^\dagger(\q_\perp,\y_\perp^\prime)\gamma^0
\big]\; .\nonumber\\
&&
\end{eqnarray}
Again, the integration over $\p_\perp$ produces a
$\delta(\y_\perp-\y_\perp^\prime)$, and we see that the matrices
${\wt U}(\y_\perp)$ and ${\wt U}^\dagger(\y_\perp^\prime)$
cancel each other to produce a unit matrix. Therefore, we conclude that:
\begin{equation}
\frac{d P_1}{d^2\q_\perp dy_q}=
\frac{d \overline{N}_{q\bar{q}}}{d^2\q_\perp dy_q}\; ,
\end{equation}
as expected.

\end{document}